\documentclass[twocolumn,showpacs,pra,superscriptaddress,longbibliography]{revtex4-1}
\usepackage{times}
\usepackage{amsmath}
\usepackage{amssymb}
\usepackage{graphicx}
\usepackage[unicode]{hyperref}
\hypersetup{
   unicode=true,          % non-Latin characters in AcrobatÕs bookmarks
   plainpages=false,
   colorlinks=true,       % false: boxed links; true: colored links
   citecolor=blue,        % color of links to bibliography
}
\usepackage[caption=false,position=top,singlelinecheck=off,justification=raggedright]{subfig}
\usepackage{color}
\usepackage{pst-node}

\def \b1{{\bf 1}}

\newcommand{\bea}{\begin{eqnarray}}

\newcommand{\eea}{\end{eqnarray}}

\newcommand{\beq}{\begin{equation}}

\newcommand{\eeq}{\end{equation}}

\begin{document}

\title{Dynamical Control of Order in a Cavity-BEC system}

\author{Jayson G. Cosme}
\affiliation{Zentrum f\"ur Optische Quantentechnologien, 
Universit\"at Hamburg, 22761 Hamburg, Germany}
\affiliation{Institut f\"ur Laserphysik, Universit\"at Hamburg, 22761 Hamburg, Germany}
\affiliation{The Hamburg Center for Ultrafast Imaging, Luruper Chaussee 149, Hamburg 22761, Germany}

\author{Christoph Georges}
\affiliation{Zentrum f\"ur Optische Quantentechnologien, 
Universit\"at Hamburg, 22761 Hamburg, Germany}
\affiliation{Institut f\"ur Laserphysik, Universit\"at Hamburg, 22761 Hamburg, Germany}

\author{Andreas Hemmerich}
\affiliation{Zentrum f\"ur Optische Quantentechnologien, 
Universit\"at Hamburg, 22761 Hamburg, Germany}
\affiliation{Institut f\"ur Laserphysik, Universit\"at Hamburg, 22761 Hamburg, Germany}
\affiliation{The Hamburg Center for Ultrafast Imaging, Luruper Chaussee 149, Hamburg 22761, Germany}

\author{Ludwig Mathey}
\affiliation{Zentrum f\"ur Optische Quantentechnologien, 
Universit\"at Hamburg, 22761 Hamburg, Germany}
\affiliation{Institut f\"ur Laserphysik, Universit\"at Hamburg, 22761 Hamburg, Germany}
\affiliation{The Hamburg Center for Ultrafast Imaging, Luruper Chaussee 149, Hamburg 22761, Germany}

%\pacs{05.70.Ln, 67.85.-d, 05.45.−a, 03.65.Sq}

\date{\today}
\begin{abstract}
We demonstrate dynamical control of the superradiant transition of cavity-BEC system via periodic driving of the pump laser. We show that the dominant density wave order of the superradiant state can be suppressed, and that the subdominant competing order of Bose-Einstein condensation emerges in the steady state. Furthermore, we show that additional, nonequilibrium density wave orders, which do not exist in equilibrium, can be stabilized dynamically. Finally, for strong driving, chaotic dynamics emerge. 
\end{abstract}
\maketitle

Recent developments in pump-probe experiments in the ultrafast regime have resulted in spectacular observations, most notably a dynamical enhancement of optical conductivity in high-$T_c$ materials, suggesting photoinduced superconductivity. This observation has been made in different materials and parameter regimes,  which leads to the question if one or more mechanisms are involved in these findings. One of the observations was reported in Ref.~\cite{Fausti2011} on pump-probe experiments in La$_{1.8-x}$Eu$_{0.2}$Sr$_x$CuO$_4$ (LESCO) at $x=1/8$ doping. Here the equilibrium material is in a charge density ordered state that strongly suppresses the superconducting dome near this commensurate doping. However, when the pump pulse is applied the superconducting response is restored. An intriguing hypothesis to explain this observation is that the pump pulse dynamically suppresses the dominant charge density wave (CDW) order allowing the next-to-leading order, i.e. superconductivity, to emerge.

\begin{figure}[!htb]
\centering
\includegraphics[width=1.0\columnwidth]{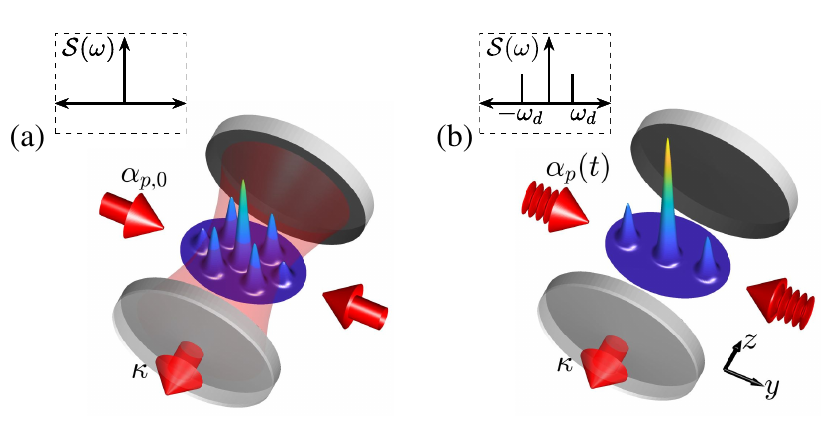}
\vspace{-1.0cm}
\caption{(a) For a transverse pump strength $\alpha_{p,0}$ above criticality, the system is in a DW phase. Atoms occupy the corresponding higher momentum states,  and photons occupy the cavity mode. (b) By modulating the pump strength $\alpha_p(t)$, DW order is suppressed and condensation is restored. The condensate density increases, and the cavity mode population is suppressed. 
 We modulate the pump beam by adding frequency sidebands $\pm\omega_d$, seen in the power spectrum $\mathcal{S}(\omega)$.}
\label{fig:summary} 
\end{figure}

We propose to test the principle of this mechanism. As a well-controlled and tunable environment \cite{Bloch2008}, we consider a cavity-Bose-Einstein condensate (BEC) system illuminated by a transverse laser beam \cite{Ritsch2013,Baumann2010,Klinder2015}. 
 As the intensity of the transverse laser beam is increased, the system undergoes a superradiant phase transition, at which the atoms self-organize into a density wave (DW) order, shown in Fig. \ref{fig:summary}(a). This DW serves as a Bragg lattice that scatters photons out of the pump laser into the cavity mode. This phase transition is related to the superradiant transition of the Dicke model \cite{Dicke1954,Hepp1973}. Two experiments, performed in different parameter regimes, have observed this transition \cite{Baumann2010,Klinder2015}. 
 Theoretical \cite{Domokos2002,Nagy2008,Bakhtiari2015,Mivehvar2017,Mivehvar2018,Gong2018} and  experimental studies \cite{Black2003,Klinder2015b,Klinder2016,Landig2016,Leonard2017} on this system have been reported. At the transition, the condensate fraction of the atomic cloud drops sharply, due to the onset of the competing density order. The phase transition displays a qualitative similarity to the competition of charge density order and superconductivity in LESCO, where  condensation is the analogue of superconducting order, and each of these orders competes with a density order. 

In this Letter, we demonstrate dynamical control of this phase transition. We show that periodic driving of the pump beam suppresses density order, and that condensation is restored, in parallel to the emergence of superconductivity due to the suppression of density order. We perform a high-frequency expansion of the Hamiltonian that demonstrates a reduction in the atom-cavity coupling parameter due to the modulation of the pump field that agrees with the numerical observation.
  The pump field modulation is realized by adding laser beams that are detuned from the pump beam. We emphasize that this choice of implementing the modulation leaves the magnitude of the pump laser unchanged so that the resulting control of the phase transition is purely dynamical.   
    Furthermore, we show that nonequilibrium DW orders arise if the driving frequency is near a resonance of the frequencies of the corresponding atomic momentum states. Finally, we observe the emergence of chaotic dynamics for strong driving.

\begin{figure}[!htb]
\centering
\includegraphics[width=1.0\columnwidth]{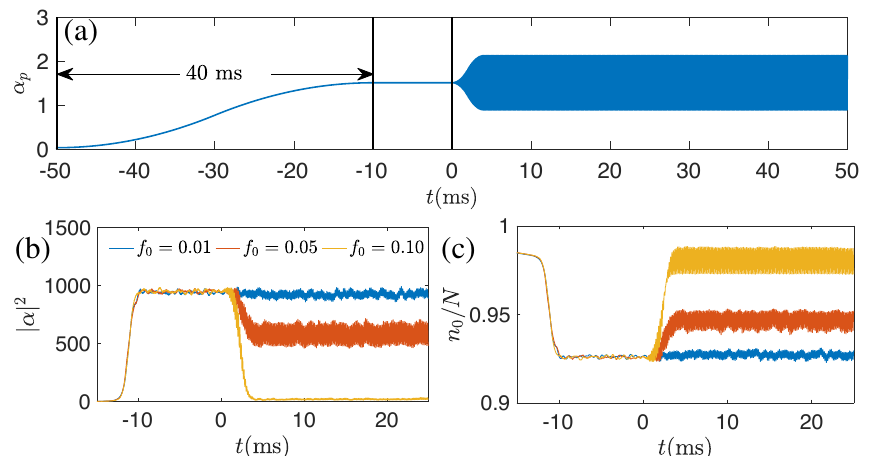}
\vspace{-0.75cm}
\caption{(a) Protocol for the pump field amplitude. Dynamics of the (b) cavity mode and (c) BEC mode occupations for $\varepsilon_0/E_{\mathrm{rec}} = 2.20$, $\omega_d = 2\pi \times 6~\mathrm{kHz}$, and different strengths of the driving amplitude $f_0$. Because of periodic driving, the density order of the atoms is dynamically suppressed, and condensation is restored.}
\label{fig:mf_dynamics} 
\end{figure}
In Fig.~\ref{fig:summary}, we depict the cavity system, with the pump laser along the $y$ direction and the cavity axis along the $z$ direction. In the rotating frame \cite{Ritsch2013}, we decompose the atomic field into plane waves ${e}^{i n k y}{e}^{i m k z}$, which gives 
\begin{align}\label{eq:ham}
&\hat{H}=-\delta_{\mathrm{C}}\hat{\alpha}^{\dagger}\hat{\alpha}+\frac{\Delta_0}{4}\hat{\alpha}^{\dagger}\hat{\alpha}\hat{Z} +\frac{\Delta_0}{2}\hat{\alpha}^{\dagger}\hat{\alpha}\hat{N} +\omega_{\mathrm{rec}}\hat{E}\\ \nonumber
& -\frac{\omega_{\mathrm{rec}}}{2}|\alpha_p|^{2}\hat{N} - \frac{\omega_{\mathrm{rec}}}{4}|\alpha_p|^{2}\hat{Y}+\frac{\sqrt{\omega_{\mathrm{rec}}|\Delta_0|}}{4}|\alpha_p|\hat{D}\hat{J}.
\end{align}
The number of atoms is $
\hat{N}=\sum \hat{\phi}^{\dagger}_{n,m}\hat{\phi}_{n,m}$ and the kinetic energy is $\hat{E}=\sum (n^2+m^2)\hat{\phi}^{\dagger}_{n,m}\hat{\phi}_{n,m}$. Momentum excitation due to the transverse pump is $\hat{Y}=\sum \left( \hat{\phi}^{\dagger}_{n+2,m}\hat{\phi}_{n,m} + \mathrm{H.c.}\right)$. The scattering of photons between the pump and the cavity fields is captured by $\hat{D}=\hat{\alpha}^{\dagger}+\hat{\alpha}$ with the momentum excitation paths $\hat{J}=\sum \left( \hat{\phi}^{\dagger}_{n,m}\left(\hat{\phi}_{n+1,m+1} + \hat{\phi}_{n+1,m-1} \right) + \mathrm{H.c.}\right)$. Excitation due to absorption and emission of cavity photons is $\hat{Z}=\sum \left( \hat{\phi}^{\dagger}_{n,m+2}\hat{\phi}_{n,m} + \mathrm{H.c.}\right)$.
 $\Delta_0$ is the light shift per intracavity photon, $\delta_C$ is the detuning between the pump and the cavity frequency, $\hat{\phi}_{n,m}$ ($\hat{\phi}^{\dagger}_{n,m}$) is the bosonic annihilation (creation) operator of the atomic momentum state $(n,m)\hbar k$, $\hat{\alpha}$ ($\hat{\alpha}^{\dagger}$) is the cavity mode  annihilation (creation) operator, and $\alpha_p$ is the dimensionless pump strength parameter \cite{Cosme2018supp}. We only consider negative detuning $\delta_{\mathrm{eff}}\equiv \delta_C - (1/2)N_a\Delta_0<0$. Photons leak out of the cavity at the rate $\kappa$.
 We use $N_a = 60\times 10^3$ atoms, $\omega_{\mathrm{rec}}=2\pi \times 3.55~\mathrm{kHz}$, $\kappa=2\pi \times 4.50~\mathrm{kHz}$, $\Delta_0= - 2\pi \times 0.36~\mathrm{Hz}$, and $\delta_{\mathrm{eff}}=-2 \pi \times 22~\mathrm{kHz}$ from \cite{Klinder2015}.

To elaborate on the analogy to high-$T_{c}$ materials, we consider the universal action of this system, to lowest order, analogous to \cite{Hayward2014,Achkar2016}. The order parameter of condensation is $\Psi = \phi_{0,0}$, the DW order parameter is $\Phi_{a} = \phi^{*}_{0,0}(\phi_{1,1} + \phi_{1,-1} +\phi_{-1,1} + \phi_{-1,-1}) + \mathrm{c.c.}$. We include the photon field as $\Phi_{ph} = \alpha$. Including only the lowest momenta and nonlinear terms, the free energy is 
 $F \approx s_{1}|\Psi|^{2} + s_{2} \Phi_{a}^{2} +s_{3} |\Phi_{ph}|^{2}+\nu_{1} |\Psi|^{2}|\Phi_{ph}|^{2}+\nu_{2} \Phi_{ph,r} \Phi_{a}$, 
 with $s_{1} = - \omega_{\mathrm{rec}} |\alpha_{p}|^{2}/2$, $s_{2} = \omega_{\mathrm{rec}}$, $s_{3}= - \omega_{\mathrm{rec}}$, $\nu_{1} = \Delta_{0}/2$, $\nu_{2} = \sqrt{\omega_{\mathrm{rec}} |\Delta_{0}|}|\alpha_{p}|/2$, and $\Phi_{ph,r}=\Re{\Phi_{ph}}$. This describes a superconducting order competing with  commensurate, real-valued DW order, where the atomic and photonic component of the DW have been treated explicitly.
 The symmetry of the system is $\mathrm{U}(1)\times \mathbb{Z}_{2}$, where the $\mathrm{U}(1)$ symmetry refers the phase invariance $\Psi \rightarrow \exp(i \theta)\Psi$, and the $\mathbb{Z}_{2}$ corresponds to the simultaneous mapping $\Phi_{ph} \rightarrow -\Phi_{ph}$ and $\Phi_{a} \rightarrow -\Phi_{a}$. 
  If the photonic mode could be integrated out without retardation we have $\Phi_{ph}\approx \nu_{2}\Phi_{a}/\delta_{C}$, so that $F_{\mathrm{eff}} \approx s_{1}|\Psi|^{2} + s'_{2} \Phi_{a}^{2} + \nu |\Psi|^{2}\Phi_{a}^{2}$, with $s'_{2} = s_{2} + s_{3}\nu_{2}^{2}/\delta_{C}^{2} + \nu_{2}^{2}/\delta_{C}$ and $\nu = \nu_{1} \nu_{2}^{2}\delta_{C}^{2}$, which shows  the competition between the BEC and DW explicitly, cf. \cite{Hayward2014}. We note, however, that the photonic dynamics cannot be integrated out without retardation. The cavity-BEC system is therefore a zero-dimensional analogue of the action in \cite{Hayward2014}, but it explicitly includes the two components of the DW order, the photonic and the atomic part.  
  
  We determine the dynamics with a numerical implementation of an open system truncated Wigner (TW) approximation \cite{Blakie2008,Polkovnikov2010}. For the initialization we choose $\alpha_{p}=0$, and we sample the initial state from a Wigner distribution of a coherent state for the BEC mode, with $\langle \phi_{0,0}\rangle  = \sqrt{N_{a}}$, and vacuum noise in all other atomic modes and the photonic mode. We propagate an initial state according to a stochastic differential equation. The unitary evolution derives from Eq.~\eqref{eq:ham}. We include a white noise $\xi(t)$, with $\langle \xi^{*}(t) \xi(t') \rangle = \kappa \delta(t-t')$ to treat photon loss to a vacuum reservoir. We use 500 trajectories to sample the dynamics, and we include momentum modes up to $\{n,m\}\in [-6,6]$. We ramp up the driving field with a protocol, shown in Fig. \ref{fig:mf_dynamics}(a). 
 We modulate the pump field $\alpha_p$ by introducing frequency sidebands $\pm\omega_d$ detuned from the pump beam 
\begin{equation}\label{eq:driving}
\alpha_p(t) = \sqrt{\epsilon_0}\left(1+f_0\mathrm{cos}{(\omega_d t)}\right),
\end{equation}
where $f_0$ is a dimensionless driving amplitude, see also Fig. \ref{fig:summary}. We emphasize that this method of driving keeps the population of the carrier frequency constant. If one would  modulate the intensity $|\alpha_p(t)|^{2}$, rather than the field $\alpha_p(t)$, there would be an additional trivial suppression of the DW phase because the intensity of the carrier frequency is decreased. Experimentally, this modulation can be achieved  by adding additional beams at frequencies that are detuned from the pump beam by $\pm \omega_d$. A version with a single frequency sideband is currently realized in \cite{Georges2018}. In Figs.~\ref{fig:mf_dynamics}(b) and \ref{fig:mf_dynamics}2(c) we show the cavity photon intensity and the BEC mode occupation as a function of time. After the ramp-up of the pump intensity, the system is in the DW phase, in which a sizable occupation of the cavity mode exists. When the modulation is turned on, the system relaxes to a steady state. As a crucial observation, we find that the coherent state is restored for driving amplitudes of $f_{0} \approx 0.1$, for this example. 

\begin{figure}[!htb]
\centering
 \includegraphics[width=0.95\columnwidth]{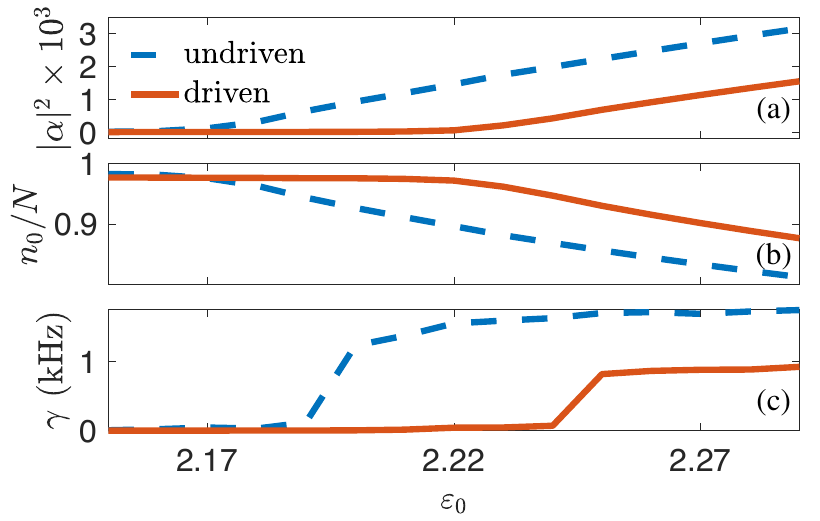}
\vspace{-0.5cm}
\caption{Comparison between the undriven and driven steady-state of (a) the cavity and (b) the BEC mode occupations and (c) the coherence decay rate as a function of the pump strength $\varepsilon_0$ (units of $E_{\mathrm{rec}}$). }
\label{fig:shift} 
\end{figure}

\begin{figure}[!ht]
\centering
\includegraphics[width=1.0\columnwidth]{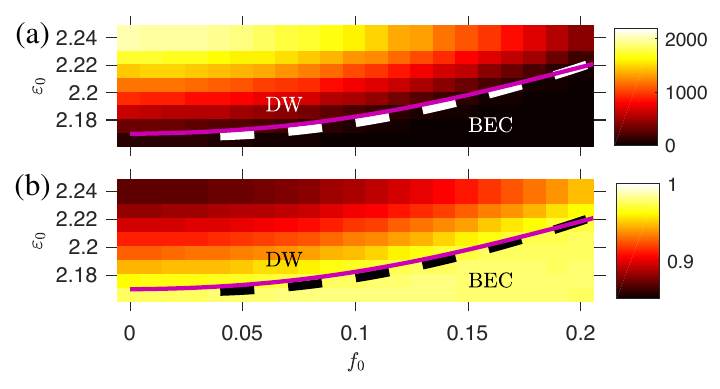}
\vspace{-0.5cm}
\caption{Dynamical renormalization of the BEC-DW phase transition, visible in the (a) cavity mode  and (b) BEC mode occupation for $w_d=2\pi\times 10~\mathrm{kHz}$. (i) Thin solid line shows the effective Hamiltonian prediction for the phase boundary, (ii) thick dashed line the TW result. The phase boundary is indicated based on $|\alpha|^2> 70$ and $n_0/N> 0.97$.}
\label{fig:mf_phase} 
\end{figure}

Next, we vary the carrier intensity $\varepsilon_0$ for fixed driving frequency $\omega_d = 2\pi \times 10~\mathrm{kHz}$ and driving amplitude $f_0=0.20$, see Fig.~\ref{fig:shift}. Panel (a) and (b) show the  cavity photon intensity and the BEC mode occupation, respectively, in the undriven state and the driven steady state. We observe that the transition from the BEC to the DW phase is shifted to a larger value of $\varepsilon_{0}$, which demonstrates dynamical control of the phase transition. 
In addition, we show the temporal correlation decay rate of the BEC mode which we determine by fitting $\langle \hat{\phi}^{\dagger}_{0,0}(t_2)\hat{\phi}_{0,0}(t_1) \rangle$ with $\sim \mathrm{exp}(-\gamma t)$.  The regime of small $\gamma$ is also extended to larger  $\varepsilon_{0}$, which demonstrates that  coherence in the BEC mode is restored.

In Fig.~\ref{fig:mf_phase}, we vary both  $\varepsilon_0$ and the driving amplitude $f_0$. The phase boundary between the BEC and DW phase is shifted to higher $\varepsilon_0$ with increasing $f_{0}$. 
We compare the numerical result to a Magnus expansion \cite{Hemmerich2010,Goldman2014,Eckardt2015,Bukov2015,Zhu2016} of the time-dependent Hamiltonian Eq.~\eqref{eq:ham}, at second order in $f_{0}$, which gives \cite{Cosme2018supp}
\begin{align}\label{eq:hameff}
&\hat{H}_\mathrm{eff}=-\delta_{\mathrm{C}}\hat{\alpha}^{\dagger}\hat{\alpha}+\frac{\Delta_0}{4}\hat{\alpha}^{\dagger}\hat{\alpha}\hat{Z} +\frac{\Delta_0}{2}\hat{\alpha}^{\dagger}\hat{\alpha}\hat{N} +\omega_{\mathrm{rec}}\hat{E}\\ \nonumber
& -\frac{\omega_{\mathrm{rec}}\epsilon_0}{2}\left(\hat{N}+\frac{f_0^2\hat{N}}{2}+\frac{\Delta_0\hat{N}}{2\omega_{\mathrm{rec}}}\left(\frac{\omega_{\mathrm{rec}}}{\omega_d}\right)^2f_0^2(\hat{D})^2 +\frac{\hat{Y}}{2} \right) \\ \nonumber
&- \frac{\omega_{\mathrm{rec}}\epsilon_0 f_0^2}{8}\hat{Y} +\frac{\sqrt{\omega_{\mathrm{rec}}|\Delta_0|\epsilon_0}}{4}\hat{D}\hat{J} \left[1-\epsilon_0\left(\frac{\omega_{\mathrm{rec}}}{\omega_d}\right)^2f_0^2\right].
\end{align}
The phase boundary predicted by Eq.~\eqref{eq:hameff} shows good agreement with the TW result, see Fig.~\ref{fig:mf_phase}. The shift of the phase boundary is primarily due to the effective reduction of the atom-cavity coupling $\sqrt{\epsilon_0} \rightarrow \sqrt{\epsilon_0}(1-\epsilon_0 \omega^2_{\mathrm{rec}} f_0^2/ \omega_d^{2})$. A dynamical renormalization of the $c$-axis transport in high-$T_c$ superconductors has been discussed in \cite{Okamoto2016}.

\begin{widetext}

\begin{figure}[!htb]
\centering
\vspace{-0.3cm}
\includegraphics[width=1.0\columnwidth]{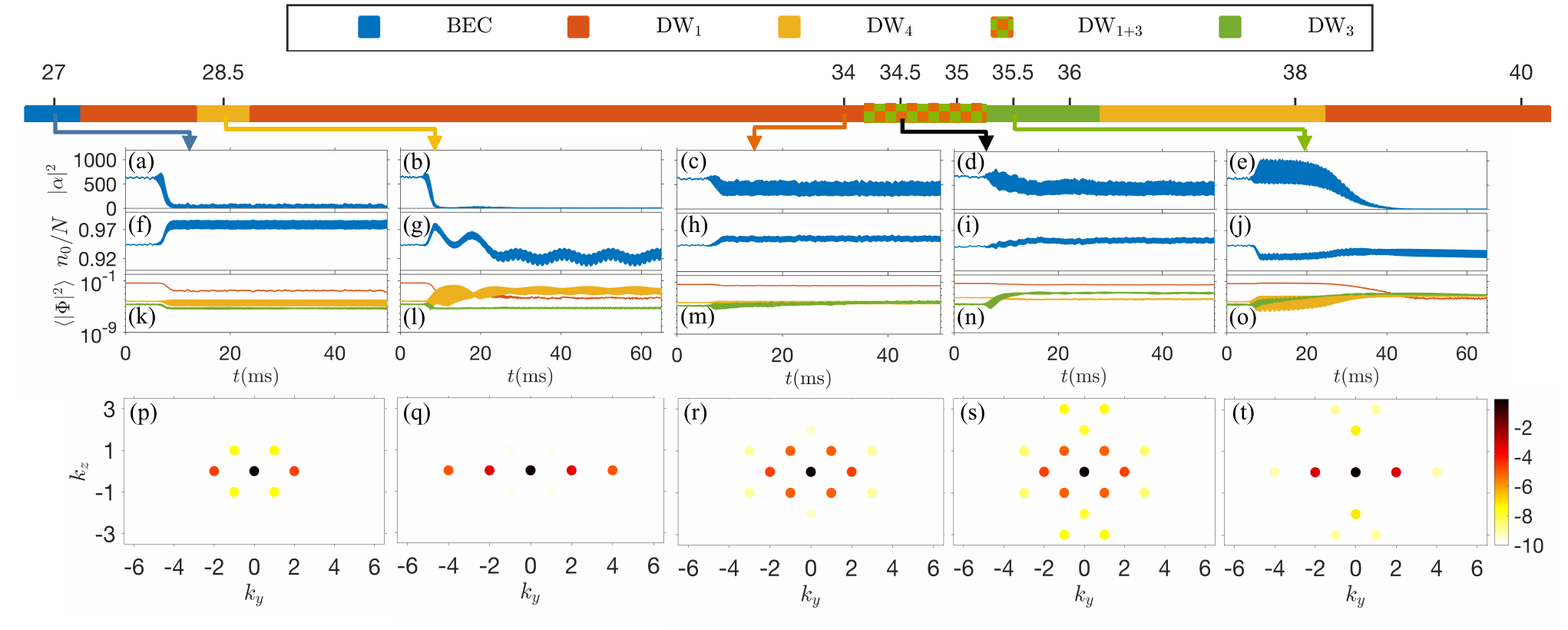}\vspace{-0.3cm}
\caption{Dynamical phase diagram as a function of $\omega_d$ in units of $2\pi\times\mathrm{kHz}$. The carrier intensity and the driving amplitude are fixed to $\varepsilon_0/E_{\mathrm{rec}} = 2.19$ and $f_0 = 0.25$, respectively. (a)-(e) Cavity and (f)-(j) BEC mode occupation dynamics, and (k)-(o) order parameter dynamics for the DW orders. For (k)-(o), each line represents the three relevant order parameters $|\Phi_{1,1}|^2$, $|\Phi_{4,0}|^2$, and $|\Phi_{1,3}|^2$. (p)-(t)  Density plot on semilogarithmic scale of the momentum occupation $|\phi_{n,m}|^2$  in the steady state, as a function of the discrete momenta $k_y$ and $k_z$.}
\label{fig:mf_morephase} 
\end{figure}
\end{widetext}

In addition to controlling the phase boundary of two equilibrium phases, we now demonstrate that we can create nonequilibrium order, see Fig.~\ref{fig:mf_morephase}.  These are orders that do not exist in equilibrium. 
In particular, we choose driving frequencies at an integer ratio to the discrete momentum $(n^2+m^2)\omega_\mathrm{rec}$, to excite new types of DW orders. In a recent work, calculations based on the Hill equation predict that parametric instabilities occur in a related system at multiples of the recoil frequency \cite{Molignini2017}.
  The associated order parameters are $\Phi_{n,m} = \mathrm{cos}(nky)\mathrm{cos}(mkz) $  as quantified by $\langle |\Phi_{n,m}|^2 \rangle$, where the DW considered above corresponds to $\Phi_{a} = \Phi_{1,1}$. We refer to this DW phase as DW$_1$.
   In addition to having a long-lived occupation of the cavity mode, the standard type of DW phase can also be identified by having a dominant order parameter given by $\Phi_{1,1}$ as seen in Fig.~\ref{fig:mf_morephase}. 
 We determine the additional higher order DW states, comparing the relative values of their order parameters.
  A new type of DW order associated with the $\phi_{\pm 4,0}$ momentum modes emerges when the driving frequency is close to half of the frequency, i.e., $2\omega_d \approx (n^2+m^2)\omega_{\mathrm{rec}} = (4^2+0^2)\omega_\mathrm{rec} = 16\omega_\mathrm{rec}$.  
  We refer to this order DW$_4$, and note that the $\phi_{\pm 4,0}$ modes are significantly occupied, in addition to the $\phi_{\pm 2,0}$ modes, as shown for $\omega_d = 2\pi \times 28.5~\mathrm{kHz}$ in Fig.~\ref{fig:mf_morephase}(q).
   Superradiance is suppressed because the condition for Bragg scattering is not fulfilled for this type of density order. This can be seen in Fig.~\ref{fig:mf_morephase}(q), where the $\phi_{\pm 1,\pm 1}$ modes are depleted for the DW$_4$ phase. Furthermore, we note that the power spectrum for the DW$_4$ phase, $\mathcal{S}_{n_0}(\omega) = |\tilde{n}_0(\omega)|^2/\int d\omega |\tilde{n}_0(\omega)|^2$, where $\tilde{n}_0(\omega)$ is the Fourier transformation of $n_0(t)$, shown in Fig.~\ref{fig:chaos_spectrum}(a), shows a subharmonic response in the dynamics of the BEC mode.  This is indicated by two prominent peaks near $\omega/\omega_d=0.5$. This is potentially related to a recent time-crystalline order proposed in Ref.~\cite{Gong2018}, but a more detailed discussion will be given elsewhere. For a driving frequency near the frequency associated with the $\phi_{\pm 1,\pm 3}$ modes, we find that its corresponding DW order, which we call DW$_3$, starts to emerge and coexist with the DW$_1$ order after transient dynamics. This intertwined order is seen for $
\omega_d =  2\pi \times 34.5~\mathrm{kHz}$ in Fig.~\ref{fig:mf_morephase}. Increasing the frequency to  $
\omega_d =  2\pi \times 35.5~\mathrm{kHz}$, we observe in Fig.~\ref{fig:mf_morephase}(e) an example for a DW$_3$ phase. Similar to the DW$_4$ phase, superradiance is suppressed as DW$_1$ order vanishes, and the order parameter for DW$_3$ becomes significant. We briefly mention that a similar emergence of metastable dynamical phases has been predicted in a periodically driven isolated Dicke model \cite{Bastidas2011}.

\begin{figure}[!htb]
\centering
\includegraphics[width=1.0\columnwidth]{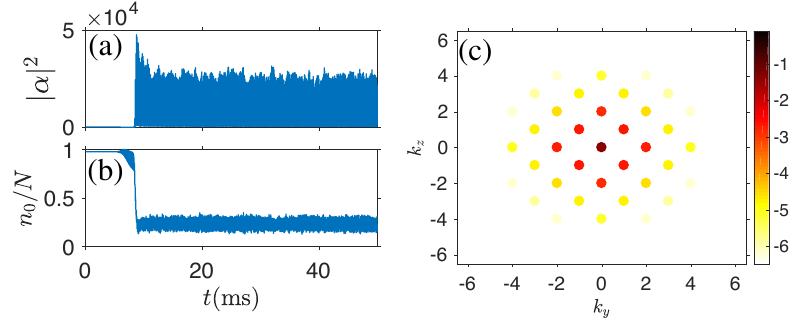}
\caption{Dynamics of (a) the cavity and (b) the BEC modes. (c) Steady state momentum occupation as in Fig.~
\ref{fig:mf_morephase}. The driving frequency is $w_d=2\pi\times 10~\mathrm{kHz}$ with $\varepsilon_0/E_{\mathrm{rec}} = 2.17$ and $f_0 = 0.90$.}
\label{fig:tw_chaos} 
\end{figure}
Finally, we show that for low driving frequency and large driving amplitude, the system enters a chaotic regime, as depicted in Fig.~\ref{fig:tw_chaos}. This phase is characterized by sharp oscillations between vanishing and the large population of the cavity mode. Because of the large cavity mode occupation, the BEC mode is severely depleted, and higher momentum modes are populated as seen in Fig.~\ref{fig:tw_chaos}. We note that, a similar dynamical phase but with regular oscillatory behavior has been discussed in \cite{Chitra2015,Molignini2017}. Here, we observe the chaotic dynamics of the observables for this chaotic phase, as seen in the power spectrum of the BEC mode dynamics and phase space trajectory presented in Fig.~\ref{fig:chaos_spectrum}.

\begin{figure}[!htb]
%\vspace{-0.5cm}
\includegraphics[width=1.0\columnwidth]{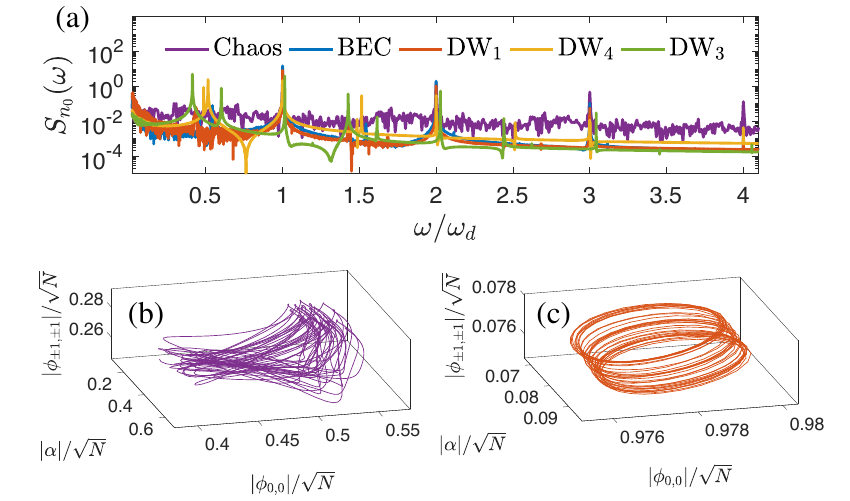}
\caption{(a) Power spectrum for the dynamics of the BEC mode for various orders shown in Figs.~\ref{fig:mf_morephase}(f-j) and \ref{fig:tw_chaos}(b). Phase space trajectory for the last 30 driving cycles for (b) chaotic regime with $f_0=0.90$ and (c) DW$_1$ with $f_0=0.25$.}
\label{fig:chaos_spectrum} 
\end{figure}

In conclusion, we have determined and characterized the dynamical states of a periodically driven cavity-BEC system. The scenario that we have described here includes the renormalization of the phase boundary of the equilibrium orders for weak to intermediate driving strengths, the emergence of nonequilibrium orders at intermediate driving strengths and at resonant driving frequencies, and chaotic dynamics for strong driving. We derive the universal action of this system which shows that it is a paradigmatic zero-dimensional system of competing orders, featuring the competition of Bose-Einstein condensation and density wave order. The density wave order itself has both an atomic and a photonic component each of which is treated explicitly. We emphasize that a broad class of many-body systems with competing orders are of this and similar form, and our study will therefore be of guidance for dynamical control in a broad, generic class of systems. 
Specifically we consider the recent finding of dynamically induced superconductivity in pump-probe experiments in the high-$T_c$ superconductor LESCO at $x=1/8$ doping. For this finding it was hypothesized that the pump pulse suppresses the CDW order, and that the subdominant order of superconductivity emerges \cite{Patel2016}. In this Letter, we have shown that the principle of this mechanism is indeed possible, and we propose it to be tested in a cavity-BEC experiment. We find that it is crucial to separate the atomic and photonic components of the DW order \cite{Cosme2018supp}, which suggests that, similarly, the electronic and atomic components of a CDW in a solid-state system have to be considered explicitly, for the emergence of nonequilibrium superconductivity, and more generally for the regime of ultrafast dynamics and the control of solid-state systems. Furthermore, the scenario that we have described beyond the renormalization of the equilibrium phase boundary, in particular nonequilibrium orders and chaotic dynamics, suggests further remarkable dynamical phenomena to be pursued in driven solid-state systems. 

\begin{acknowledgments}
We would like to acknowledge the support from the Deutsche Forschungsgemeinschaft through the SFB 925 and the Hamburg Centre for Ultrafast Imaging. We also thank Andrea Cavalleri, Louis-Paul Henry, Jun-ichi Okamoto, and Beilei Zhu for useful discussions. 
\end{acknowledgments}

\bibliography{biblio}

\newpage

\onecolumngrid
\begin{center}
\textbf{\large Supplemental Materials: Dynamical control of order in a cavity-BEC system}
\vspace{0.6cm}

Jayson G. Cosme$^{1,2,3}$, Christoph Georges$^{1,2}$, Andreas Hemmerich$^{1,2,3}$, and Ludwig Mathey$^{1,2,3}$\\

$^1${\it Zentrum f\"ur Optische Quantentechnologien, Universit\"at Hamburg, 22761 Hamburg, Germany}

$^2${\it Institut f\"ur Laserphysik, Universit\"at Hamburg, 22761 Hamburg, Germany}

$^3${\it The Hamburg Center for Ultrafast Imaging, Luruper Chaussee 149, Hamburg 22761, Germany}

\end{center}

\setcounter{equation}{0}
\setcounter{figure}{0}
\setcounter{table}{0}
\renewcommand{\theequation}{S\arabic{equation}}
\renewcommand{\thefigure}{S\arabic{figure}}

\section{Dynamical protocol}
The full dynamical protocol used in obtaining the mean-field results consists of two stages: (i) slow ramp towards the mean pump amplitude $\alpha_p=\sqrt{\epsilon_0}$; and (ii) the driving protocol. The exact time-dependence is shown below: 
\begin{align}\label{eq:real_driving}
        &\alpha_p(t)= \\ \nonumber
       &\left\{ 
        \begin{array}{ll}
        	\sqrt{\epsilon_0}B_1(t+T_\mathrm{r}+T_\mathrm{c},T_\mathrm{r}) &t \in [-T_\mathrm{r}-T_\mathrm{c}, \frac{-T_\mathrm{r}-2T_\mathrm{c}}{2}] \\
        	\sqrt{\epsilon_0}B_2(t+T_\mathrm{r}+T_\mathrm{c},T_\mathrm{r}) &t \in (\frac{-T_\mathrm{r}-2T_\mathrm{c}}{2}, -T_\mathrm{c}] \\
        	\sqrt{\epsilon_0} &t \in (-T_\mathrm{c}, 0] \\
            \sqrt{\epsilon_0}(1+B_1(t,T_\mathrm{s})f_0\mathrm{cos}(\omega_d t)) &t \in (0, T_{\mathrm{s}}/2] \\
            \sqrt{\epsilon_0}(1+B_2(t,T_\mathrm{s})f_0\mathrm{cos}(\omega_d t)) &t \in (T_{\mathrm{s}}/2, T_{\mathrm{s}}] \\
            \sqrt{\epsilon_0}(1+f_0\mathrm{cos}(\omega_d t)) &t>T_{\mathrm{s}}
        \end{array} 
        \right.
\end{align}
where
\begin{align}
B_1(t,T)&=\frac{2t^2}{T^2} \\ \nonumber
B_2(t,T)&=-1-\frac{2t^2}{T^2}+\frac{4t}{T}.
\end{align}
Specifically, we have chosen $T_\mathrm{r}=40~\mathrm{ms}$, $T_\mathrm{c}=10~\mathrm{ms}$, and $T_\mathrm{s}=4~\mathrm{ms}$. Note that the actual experimental value for the pump beam intensity $|\varepsilon_0|$ for the setup in Ref.~\cite{Klinder2015} can be modelled within the single mode description by an effective reduction in the coupling according to $|\varepsilon_0|/|\epsilon_0| \approx 1.44 E_{\mathrm{rec}} $ where $E_{\mathrm{rec}}$ is the recoil energy and $\epsilon_0$ is the dimensionless pump strength parameter used in the single-mode model.
\begin{figure}[ht!]
\includegraphics[width=0.5\columnwidth]{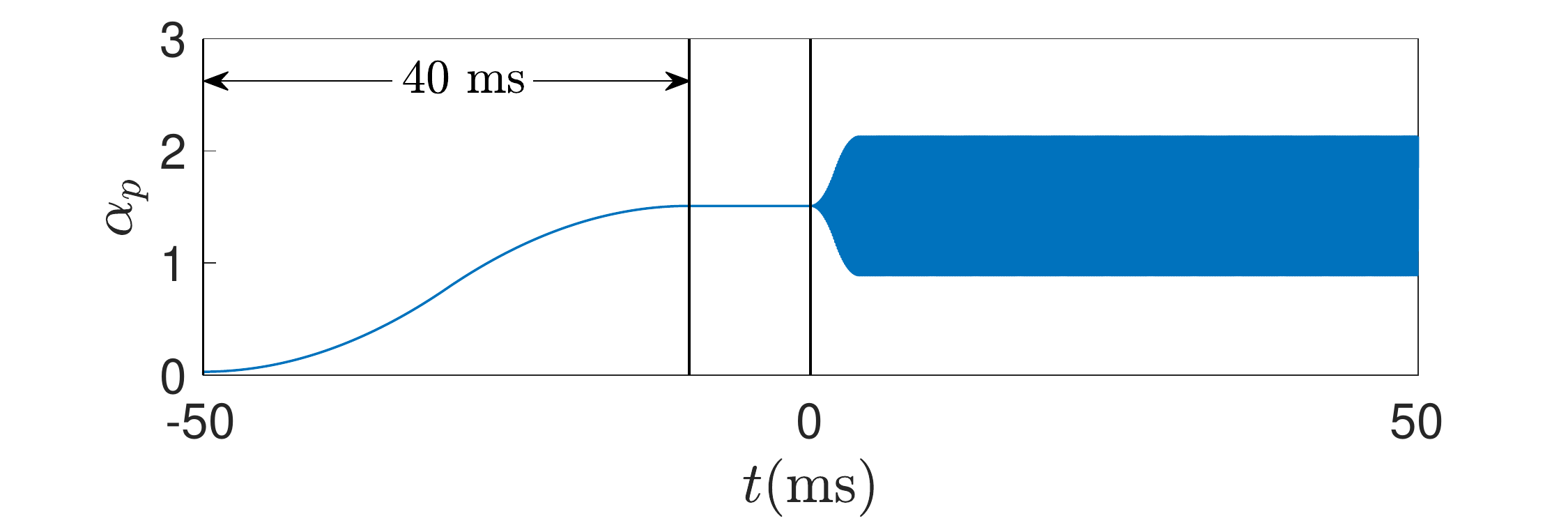}\includegraphics[width=0.5\columnwidth]{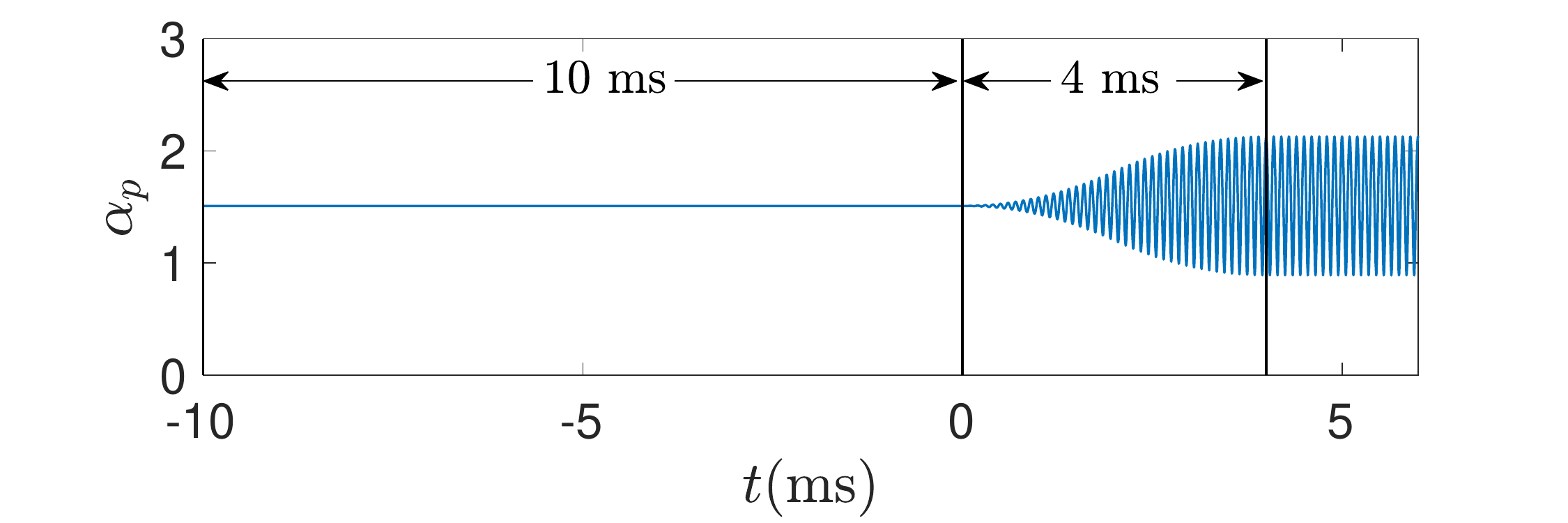}
\caption{Time evolution of the pump field amplitude $\alpha_p$.
}\label{fig:proto}
\end{figure}
A schematic for the time evolution of the pump field used in this work is shown in Fig.~\ref{fig:proto}. As seen in Fig.~\ref{fig:proto}, the system is allowed to evolve and relax for more than $50~\mathrm{ms}$ upon reaching the desired modulation strength. The long-time average of relevant observables shown in the main text correspond to a time averaging over the final $10~\mathrm{ms}$ of the full dynamics.
\begin{figure}[ht!]
\includegraphics[width=0.5\columnwidth]{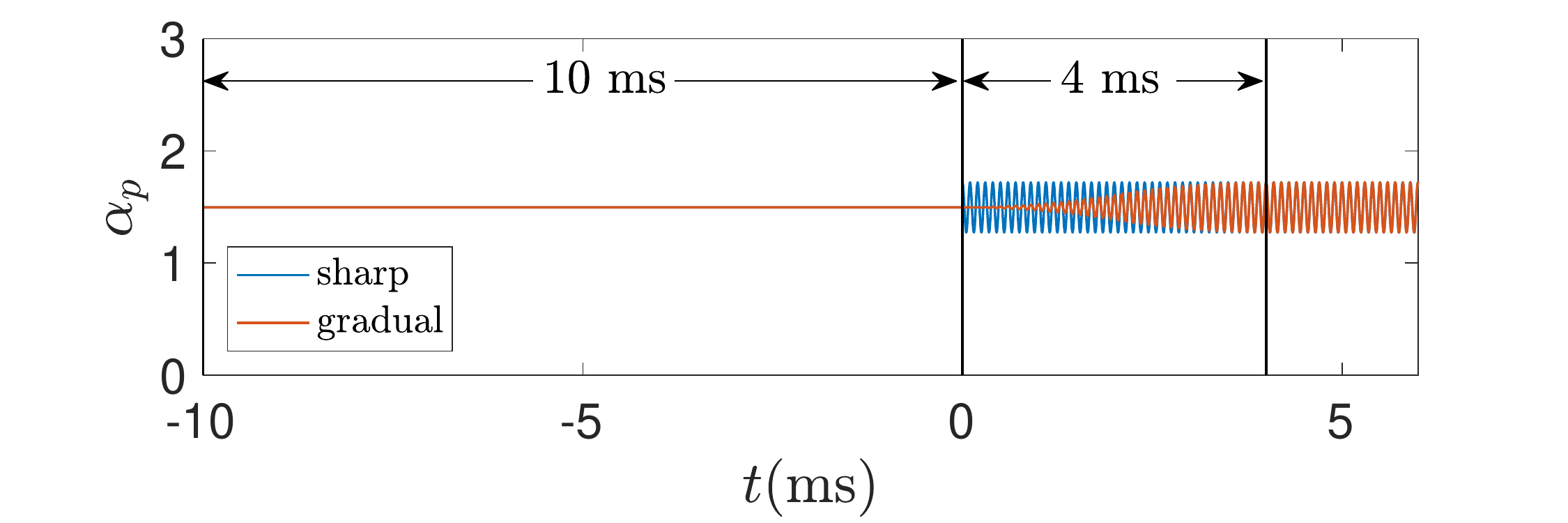}\includegraphics[width=0.5\columnwidth]{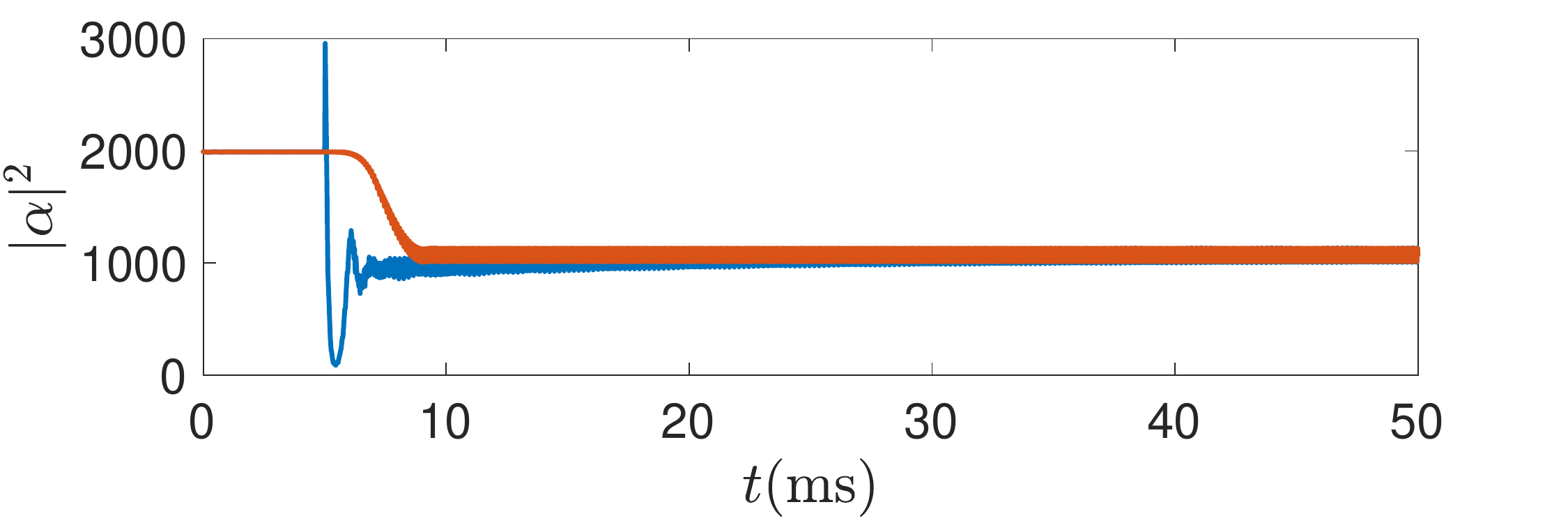}
\caption{(Left) Difference between the time evolution of the pump field amplitude $\alpha_p$ for a sharp and a gradual increase in the driving amplitude $f_0$. (Right) Comparison of the corresponding dynamcis of the cavity mode between the two driving protocols. Here, we have chosen $\varepsilon_0/E_{\mathrm{rec}} = 2.24$, $\omega_d = 2\pi \times 10~\mathrm{kHz}$, and $f_0=0.15$.
}\label{fig:protolt}
\end{figure}

We show in Fig.~\ref{fig:protolt} a comparison for the dynamical response of the system between a sharp and a gradual switching of the driving amplitude. There, it can be seen that the two protocols only differ in the dynamical response of the system for short times but the long-time average of observables is the same in both protocols.

\section{Advantage of double-sideband over single-sideband protocol}

There are two obvious ways to drive the pump intensity. The first one used in the main text is generated by introducing two additional sidebands at $\pm \omega_d$. Recall that for this case, we have 
\begin{equation}
\alpha^{\{2\}}_p(t) = \sqrt{\epsilon_0}(1 + f_0 \mathrm{cos}(\omega_d t)),
\end{equation}
and this creates an intensity modulation for the pump according to
\begin{equation}\label{eq:prot2}
|\alpha^{\{2\}}_p(t)|^2 = {\epsilon_0}\left(1 + \frac{f_0^2}{2} + \frac{f_0^2\mathrm{cos}(2\omega_d t)}{2} + 2f_0 \mathrm{cos}(\omega_d t)\right).
\end{equation}
On the other hand, a second type of driving can be realized by adding just a single sideband say for example at $+ \omega_d$. This single-sideband protocol can be expressed as
\begin{equation}
\alpha^{\{1\}}_p(t) = \sqrt{\epsilon_0}(1 + f_0 e^{i\omega_d t}),
\end{equation}
which then drives the pump beam intensity given by
\begin{equation}\label{eq:prot1}
|\alpha^{\{1\}}_p(t)|^2 = {\epsilon_0}\left(1 + {f_0^2} + 2f_0 \mathrm{cos}(\omega_d t)\right).
\end{equation}
If we compare Eqs.~\eqref{eq:prot2} and \eqref{eq:prot1}, it becomes immediately obvious that the single-sideband protocol introduces a larger constant shift of $\epsilon_0f_0^2$ to the pump power as compared to the double-sideband protocol which only increases the pump intensity by a constant amount of $\epsilon_0f_0^2/2$. This becomes problematic for larger values of $\varepsilon_0$ which require stronger driving amplitude if one intends to completely wipe the DW phase. Indeed, as shown in an example presented in Fig.~\ref{fig:proto}, the reduction in the number of cavity photons is much greater in the double-sideband protocol for a fixed value of the driving amplitude $f_0$. 
\begin{figure}[ht!]
\includegraphics[width=0.45\columnwidth]{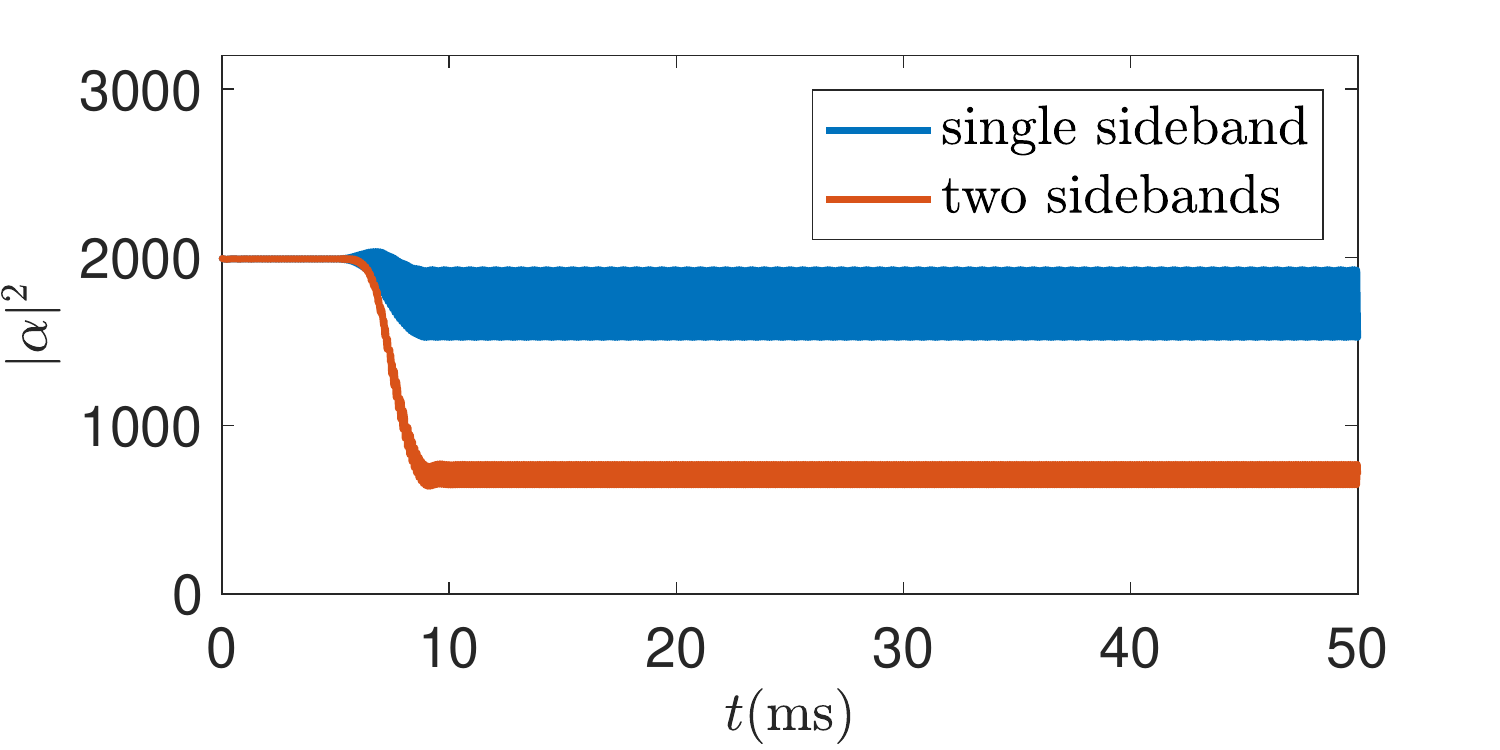}
\caption{Comparison between the suppression effect of single-sideband and double-sideband protocols. Time evolution of the cavity mode occupation for $\varepsilon_0/E_{\mathrm{rec}} = 2.24$, $\omega_d = 2\pi \times 10~\mathrm{kHz}$, and $f_0=0.18$.
}\label{fig:svd}
\end{figure}

\section{Temporal correlation}
\begin{figure}[!htb]
\centering
\includegraphics[width=0.45\columnwidth]{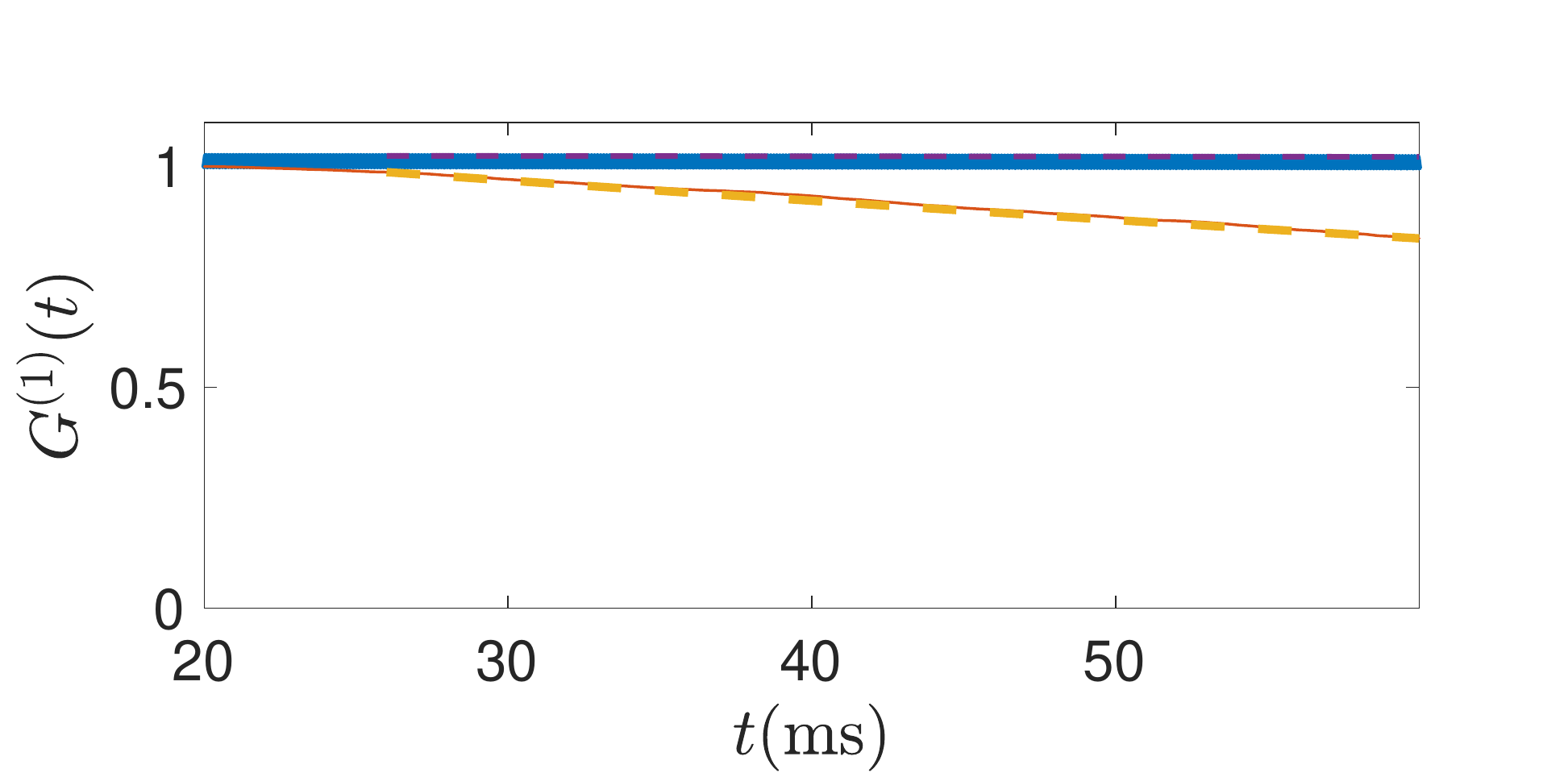}\includegraphics[width=0.45\columnwidth]{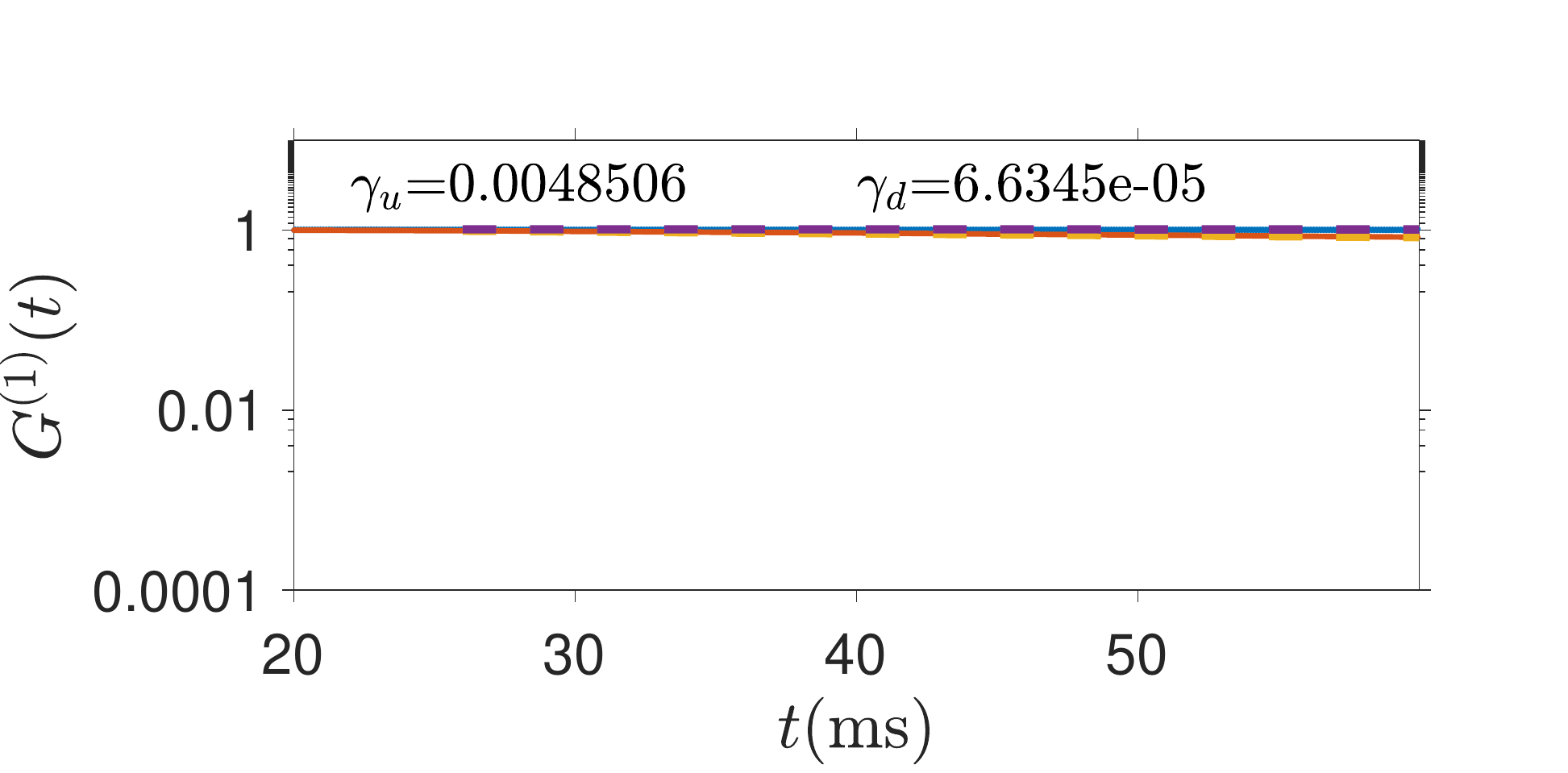}\\
\includegraphics[width=0.45\columnwidth]{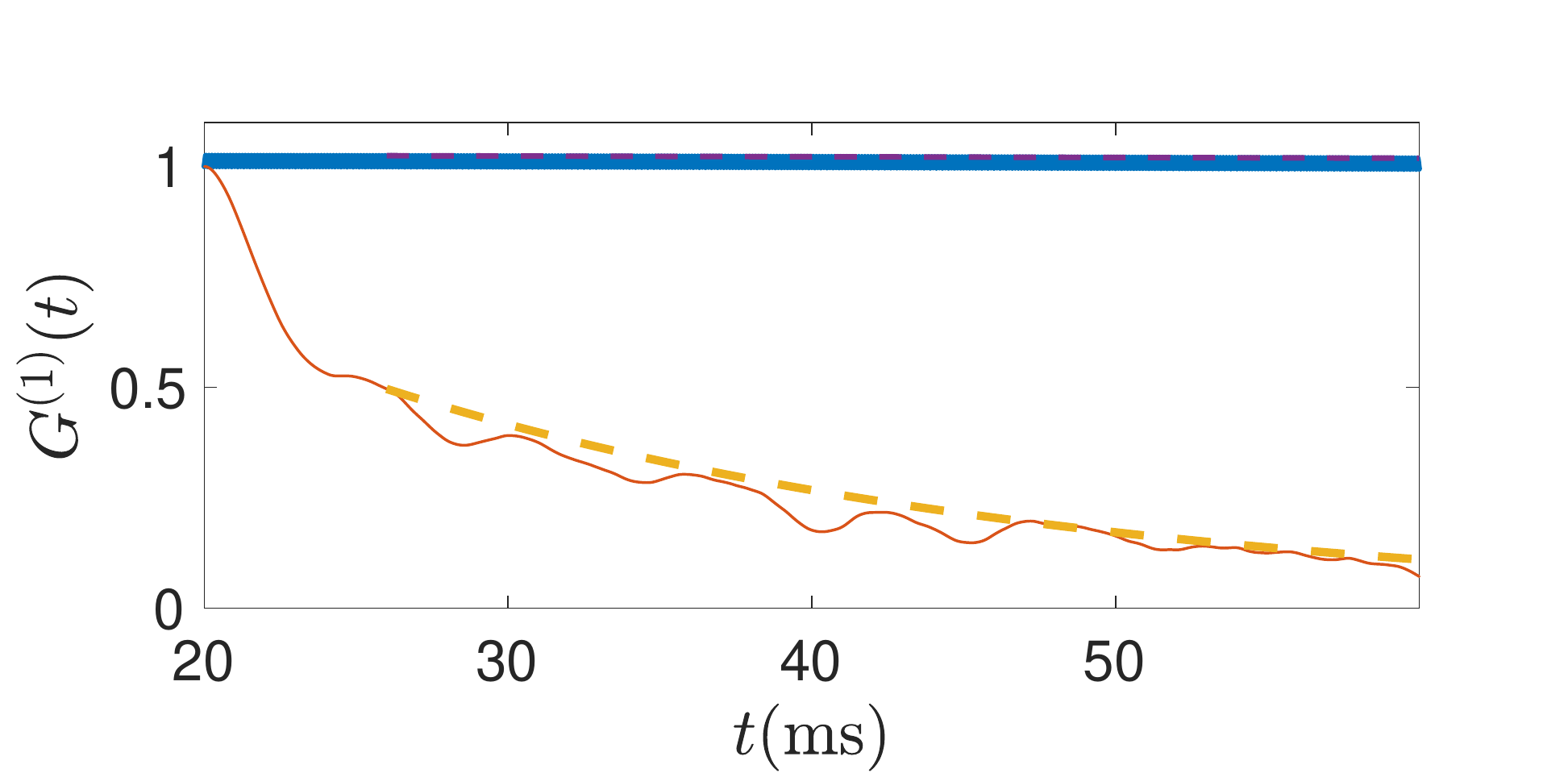}\includegraphics[width=0.45\columnwidth]{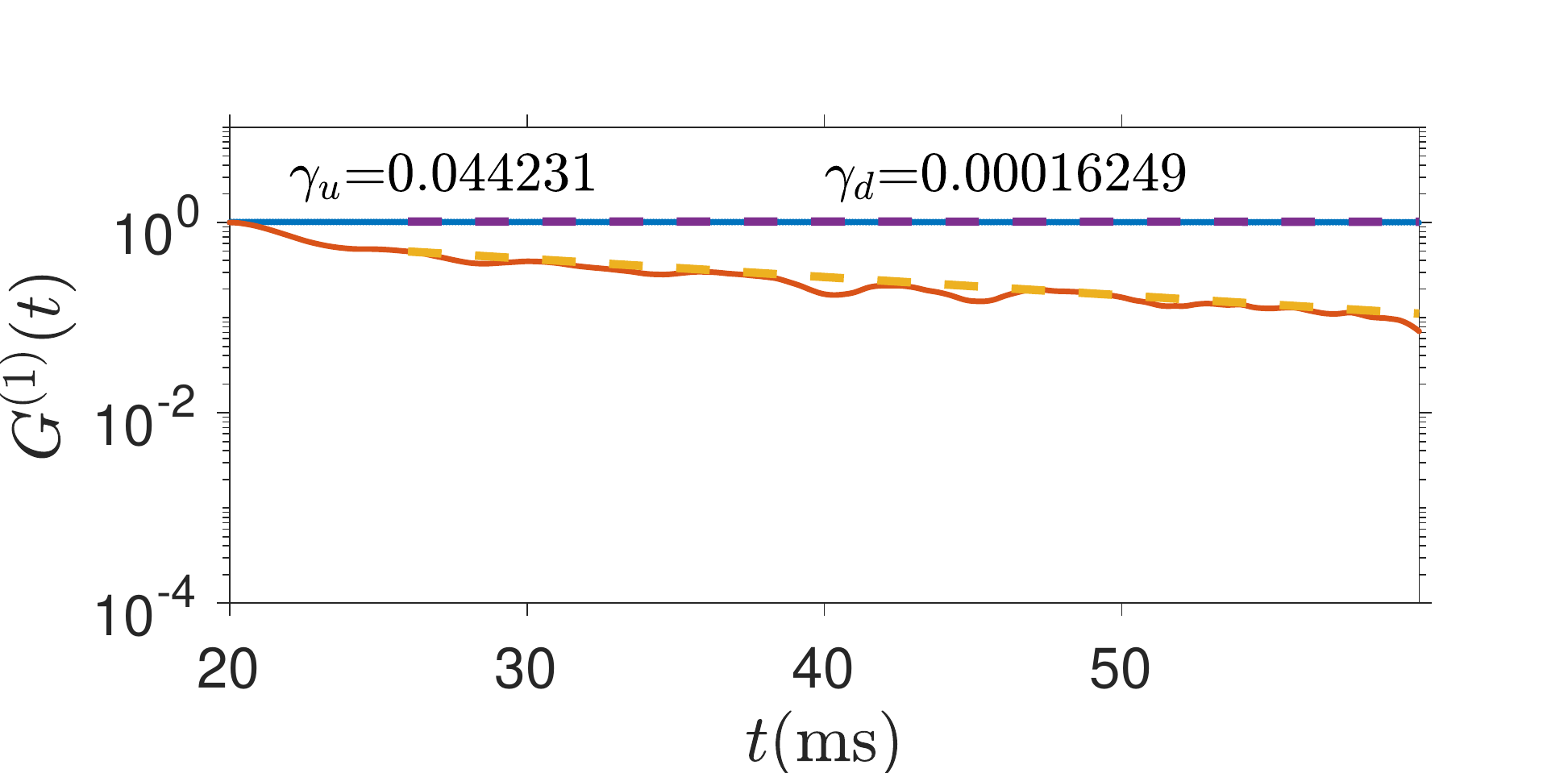}\\
\includegraphics[width=0.45\columnwidth]{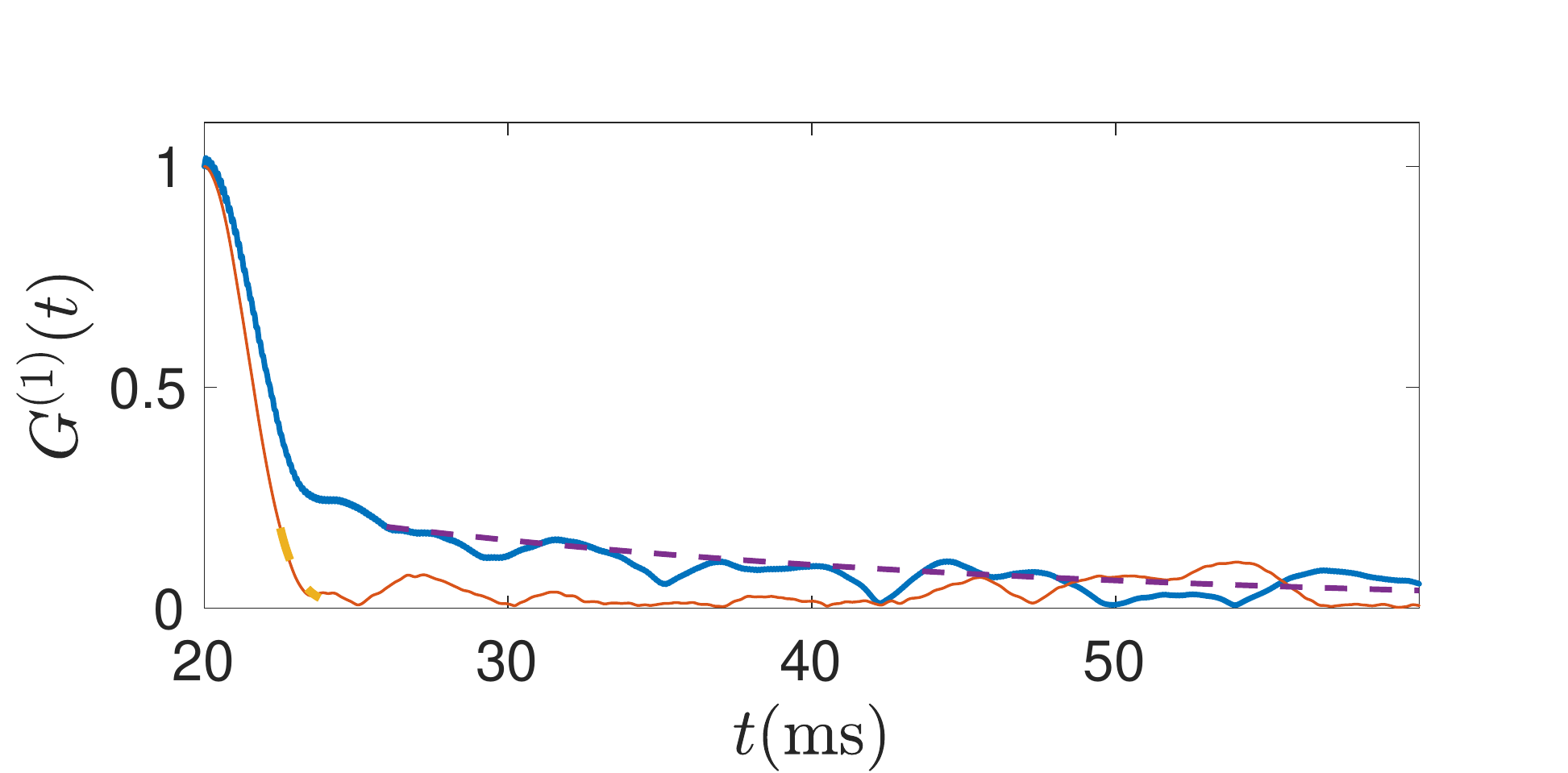}\includegraphics[width=0.45\columnwidth]{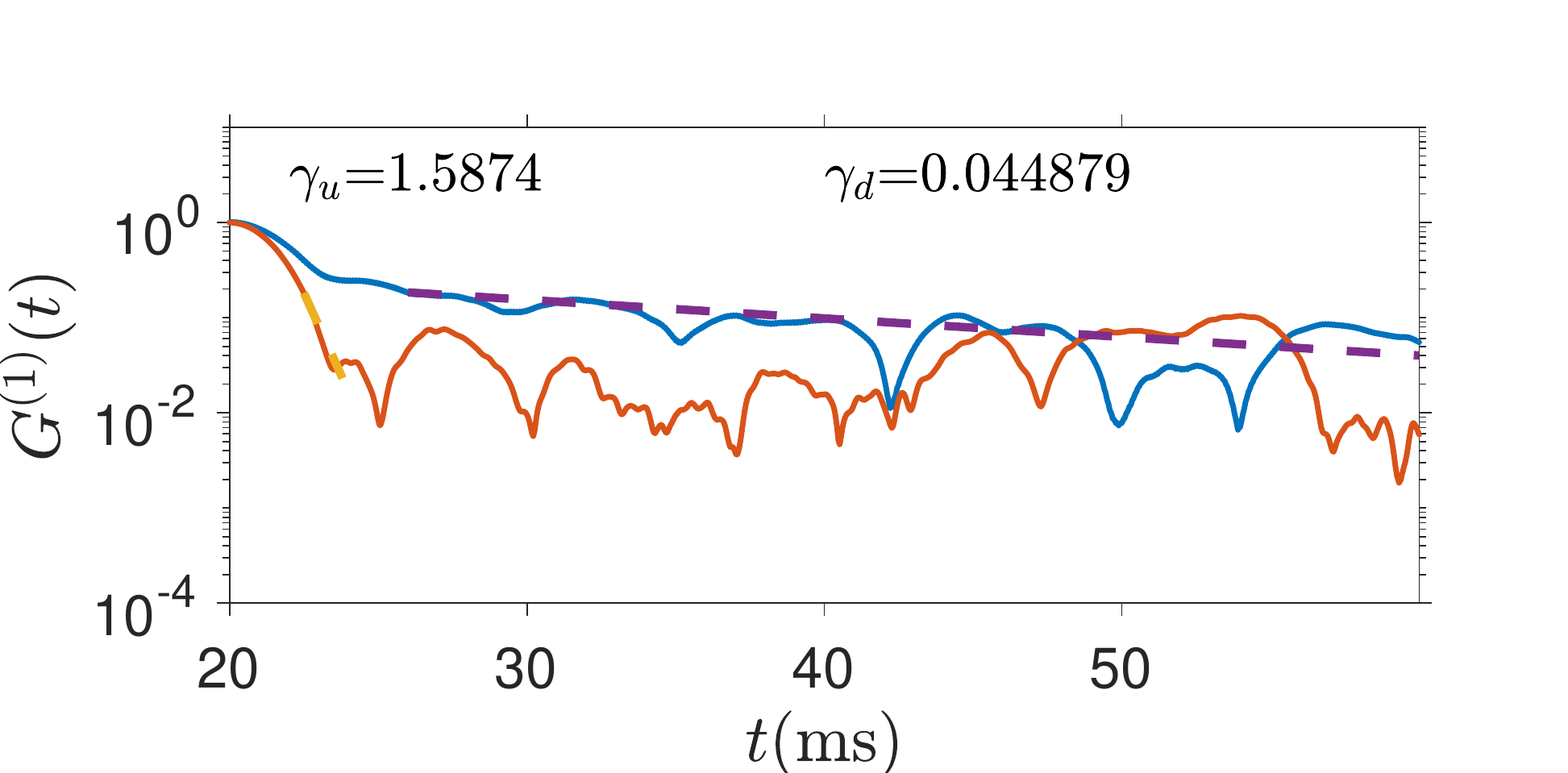}\\
\caption{Temporal correlation for the (Red) undriven case and the (blue) driven case. (Top to bottom) $\varepsilon_0 = \{2.15,2.17,2.23\}$. (Left) linear and (right) semi-logarithmic scale. Dashed curves correspond to the exponential fit as described in the text. }
\label{fig:mf_dynamics_N00} 
\end{figure}

In order to obtain the dependence of the temporal correlation on the pump strength parameter shown in Fig. 3(c), we first calculate the temporal correlation according to
\begin{equation}
G^{(1)}(t) = \frac{\left( \left(\mathrm{Re}\langle \hat{\phi}^{\dagger}_{0,0}(t)\hat{\phi}_{0,0}(t_1) \rangle\right)^2 +  \left(\mathrm{Im}\langle \hat{\phi}^{\dagger}_{0,0}(t)\hat{\phi}_{0,0}(t_1) \rangle\right)^2 \right)^{1/2}}{\langle n_{0,0}(t_1) \rangle}
\end{equation}
where $t_1=20~\mathrm{ms}$. The corresponding decay rates indicative of the correlation time in the system are $\gamma_u$ for the undriven case and $\gamma_d$ for the driven case. This can be extracted from fitting an exponential decay $\mathrm{exp}(-\gamma t)$ to $G^{(1)}(t)$ as exemplified in Fig.~\ref{fig:mf_dynamics_N00}.

\section{Comparison between mean-field and truncated Wigner results}
In order include quantum fluctuations, we have simulated the dynamics within the truncated Wigner (TW) approximation. A detailed discussion of this method and how to sample the initial quantum noise for coherent and vacuum states can be found in \cite{Blakie2008,Polkovnikov2010}. In a nutshell, the TW approximation goes beyond the mean-field level by accounting for quantum fluctuations in the initial state of the system. This is done by solving the underlying mean-field equations of motion stochastically using an ensemble of initial conditions or trajectories that correctly samples the initial Wigner distribution for the available quantum states in the system. Finally, observables obtained from each trajectory are averaged over the ensemble. For the cavity-BEC system considered in this work, the corresponding set of mean-field equation reads \cite{Ritsch2013}
\begin{align}\label{eq:eom}
	i \frac{\partial \phi_{n,m}}{\partial t} &= 
	\omega_{\mathrm{rec}}\left(n^2 + m^2+\frac{\Delta_0}{2\omega_{\mathrm{rec}}}|\alpha|^2-\frac{|\alpha_p(t)|^2}{2}\right)\phi_{n,m} +\frac{\Delta_0}{4}|\alpha|^2(\phi_{n,m-2}+\phi_{n,m+2})-\frac{\omega_{\mathrm{rec}}}{4}|\alpha_p(t)|^2(\phi_{n-2,m}+\phi_{n+2,m})\\ \nonumber
	 &+\frac{\sqrt{\omega_{\mathrm{rec}}}\sqrt{|\Delta_0|}}{2}\alpha_p(t)\mathrm{Re}({\alpha})(\phi_{n-1,m-1}+\phi_{n+1,m-1}+\phi_{n-1,m+1}+\phi_{n+1,m+1})\\ \nonumber
	 i \frac{\partial \alpha}{\partial t} &= \left[-\delta_{\mathrm{eff}} + \frac{1}{2}N_a\Delta_0 \sum_{n,m}\mathrm{Re}[\phi_{n,m}\phi^{*}_{n,m+2}]-i\kappa \right]\alpha + i\xi \\ \nonumber
	 &+\frac{N_a\sqrt{\omega_{\mathrm{rec}}}\sqrt{|\Delta_0|}}{4}\alpha_p(t)\sum_{n,m}\phi_{n,m}(\phi^{*}_{n+1,m+1}+\phi^{*}_{n+1,m-1}) + \phi^{*}_{n,m}(\phi_{n+1,m+1}+\phi_{n+1,m-1}),
\end{align}
where the Gaussian noise operator $\xi$ in the cavity mode equation follows $\langle \xi(t)\xi^{\dagger}(t') \rangle= \kappa\delta(t-t')$. 
\begin{figure}[!htb]
\centering
\includegraphics[width=0.7\columnwidth]{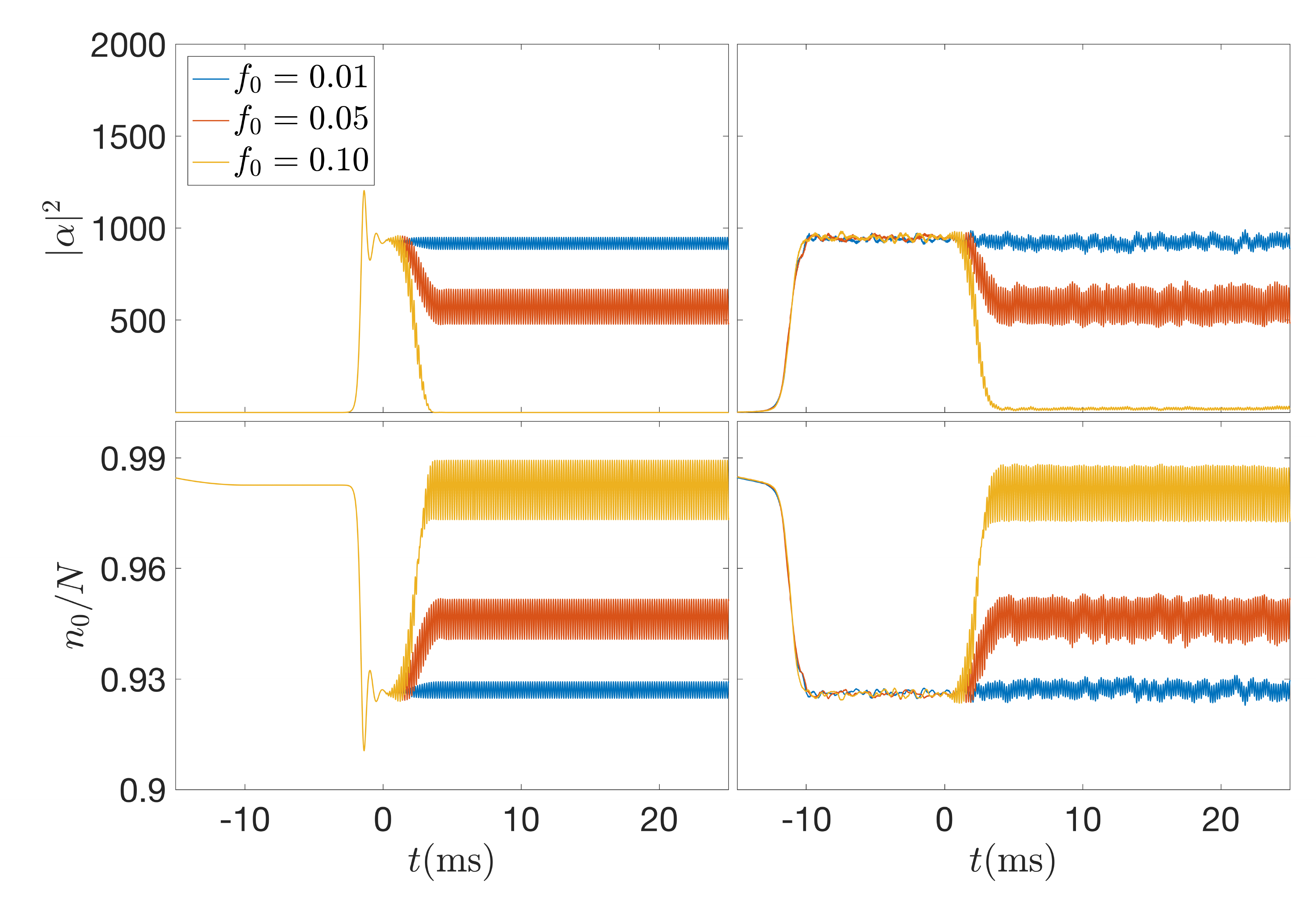}
\caption{(Left) Mean-field and (Right) truncated Wigner dynamics for the (Top) cavity mode and (Bottom) BEC mode occupations for $\varepsilon_0/E_{\mathrm{rec}} = 2.20$, $\omega_d = 2\pi \times 6~\mathrm{kHz}$, and different strengths of the driving amplitude $f_0$. }
\label{fig:mf_tw_dynamics} 
\end{figure}

A comparison between the mean-field and TW results are shown in Fig.~\ref{fig:mf_tw_dynamics}. As seen in Fig.~\ref{fig:mf_tw_dynamics}, the main difference between the mean-field and truncated Wigner simulations is the apparent earlier onset of DW formation predicted by TWA. This suggests that quantum fluctuations lower the threshold value for the phase transition from the BEC to the DW phase. A more in-depth discussion about this phenomenon and how it modifies the hysteretic dynamics observed in Ref.~\cite{Klinder2015} will be addressed in an upcoming work \cite{Cosme2018futu}. Apart from this deviation, it can be seen that the ability to dynamically control the BEC and density-ordered phases in the system appears to be robust against quantum and vacuum fluctuations from the initial state. 

\section{Derivation of the Effective Time-Independent Hamiltonian}
Recall that the Hamiltonian shown in the main text reads
\begin{align}\label{eq:hamilt}
	\hat{H}&= -\delta_{\mathrm{C}}\hat{\alpha}^{\dagger}\hat{\alpha} + \frac{\Delta_0}{4}\hat{\alpha}^{\dagger}\hat{\alpha}\sum_{n,m}\left( \hat{\phi}^{\dagger}_{n,m+2}\hat{\phi}_{n,m} +\hat{\phi}^{\dagger}_{n,m}\hat{\phi}_{n,m+2} \right) +\frac{\Delta_0}{2}\hat{\alpha}^{\dagger}\hat{\alpha}\sum_{n,m}\hat{\phi}^{\dagger}_{n,m}\hat{\phi}_{n,m} \\ \nonumber
	 &+ \omega_{\mathrm{rec}}\sum_{n,m}(n^2+m^2)\hat{\phi}^{\dagger}_{n,m}\hat{\phi}_{n,m} -\frac{\omega_{\mathrm{rec}}}{2}|\alpha_p|^{2}\sum_{n,m}\hat{\phi}^{\dagger}_{n,m}\hat{\phi}_{n,m} - \frac{\omega_{\mathrm{rec}}}{4}|\alpha_p|^{2}\sum_{n,m}\left( \hat{\phi}^{\dagger}_{n+2,m}\hat{\phi}_{n,m} +\hat{\phi}^{\dagger}_{n,m}\hat{\phi}_{n+2,m} \right) \\ \nonumber
	 &+\frac{\sqrt{\omega_{\mathrm{rec}}}\sqrt{|\Delta_0|}}{4}|\alpha_p|(\hat{\alpha}^{\dagger}+\hat{\alpha})\sum_{n,n}\left( \hat{\phi}^{\dagger}_{n,m}(\hat{\phi}_{n+1,m+1}+\hat{\phi}_{n+1,m-1}) + (\hat{\phi}^{\dagger}_{n+1,m+1}+\hat{\phi}^{\dagger}_{n+1,m-1})\hat{\phi}_{n,m} \right),
\end{align}
For the kind of driving considered here, the pump field amplitude is driven according to
\begin{equation}\label{eq:side}
\alpha_p(t) = \sqrt{\epsilon_0}(1 + f_0 \mathrm{cos}(\omega_d t)),
\end{equation}
which effectively drives the pump beam intensity via
\begin{equation}
|\alpha_p(t)|^2 = {\epsilon_0}\left(1 + \frac{f_0^2}{2} + \frac{f_0^2\mathrm{cos}(2\omega_d t)}{2} + 2f_0 \mathrm{cos}(\omega_d t)\right).
\end{equation}
An effective time-independent Hamiltonian can be obtained from Floquet-Magnus \cite{Bukov2015,Zhu2016} or high-frequency expansion \cite{Hemmerich2010,Goldman2014,Eckardt2015}. We briefly outline the general procedure for such expansion below. To this end, it is helpful to expand the time-dependent Hamiltonian in terms of its Fourier components such that
\begin{equation}
\hat{H}(t) \equiv \hat{H} = \sum_{m=-\infty}^{\infty} e^{im\omega_d t}\hat{H}_m.
\end{equation}
The effective Hamiltonian can then be expanded as
\begin{equation}
H_\mathrm{eff} = \sum_{n=0}^{\infty} H^{(n)}_{\mathrm{eff}}
\end{equation}
where up to second-order we have \cite{Goldman2014,Bukov2015,Eckardt2015}
\begin{align}
&H^{(0)}_{\mathrm{eff}} = H_0 \\ \nonumber
&H^{(1)}_{\mathrm{eff}} = \frac{1}{\omega_d}\sum_{\ell} \frac{1}{\ell}[H_{\ell},H_{-\ell}] \\ \nonumber
&H^{(2)}_{\mathrm{eff}} =  \frac{1}{\omega_d^2}\sum_{\ell \neq 0} \left(\frac{[H_{-\ell},[H_{0},H_{\ell}]]}{2\ell^2} + \sum_{\ell'\neq 0,\ell} \frac{[H_{-\ell'},[H_{\ell'-\ell},H_{\ell}]]}{3\ell\ell'} \right).
\end{align}
For a single frequency sideband as in Eq.~\eqref{eq:prot1}, the first non-trivial correction to the time-averaged Hamiltonian $H_0$ is given by the first-order correction $H^{(1)}_{\mathrm{eff}}$ since $H_1 \neq H_{-1}$ in this case. However for the two-sideband protocol considered in this work, $H_1 = H_{-1}$ meaning the first-order correction for the effective Hamiltonian is zero, $H^{(1)}_{\mathrm{eff}}=0$. Therefore, we have the following effective time-independent Hamiltonian
\begin{equation}\label{eq:2Ham}
H_\mathrm{eff} = H_0 + H^{(2)}_{\mathrm{eff}}.
\end{equation}
where
\begin{equation}\label{eq:2eff}
H^{(2)}_{\mathrm{eff}}=-\frac{1}{4\omega_d^2}\left[[H_0,A_1],A_1\right]
\end{equation}
Note that in Eq.~\eqref{eq:2eff}, we have introduced
\begin{equation}\label{eq:h0}
H_0 = -\delta_{\mathrm{C}}C + \frac{\Delta_0}{4}CZ + \omega_{\mathrm{rec}}E +\frac{\Delta_0}{2}CN -\frac{\omega_{\mathrm{rec}} \epsilon_0}{2}\left(1+\frac{f_0^2}{2}\right)N -\frac{\omega_{\mathrm{rec}} \epsilon_0}{4}\left(1+\frac{f_0^2}{2}\right)Y +\frac{\sqrt{\omega_{\mathrm{rec}}}\sqrt{|\Delta_0|} \sqrt{\epsilon_0}}{4}DJ
\end{equation}
and
\begin{equation}\label{eq:a1}
2H_1 = 2H_{-1} \equiv A_1=-\frac{\omega_{\mathrm{rec}} (2f_0\epsilon_0)}{2}N -\frac{\omega_{\mathrm{rec}} (2f_0\epsilon_0)}{4}Y +\frac{\sqrt{\omega_{\mathrm{rec}}} \sqrt{|\Delta_0|}\sqrt{\epsilon_0} f_0}{4 }DJ.
\end{equation}
For brevity we will drop the hats in the operators. Note that in Eqs~\eqref{eq:h0} and \eqref{eq:a1}, we define the following operators:
\begin{align}
C&=\alpha^{\dagger}\alpha \\ \nonumber
D&=\alpha^{\dagger}+\alpha \\ \nonumber
N&=\sum \phi^{\dagger}_{n,m}\phi_{n,m} \\ \nonumber
E&=\sum (n^2+m^2)\phi^{\dagger}_{n,m}\phi_{n,m} \\ \nonumber
Z&=\sum \left( \phi^{\dagger}_{n,m+2}\phi_{n,m} + \mathrm{h.c.}\right)\\ \nonumber
Y&=\sum \left( \phi^{\dagger}_{n+2,m}\phi_{n,m} + \mathrm{h.c.}\right)\\ \nonumber
J&=\sum \left( \phi^{\dagger}_{n,m}\left(\phi_{n+1,m+1} + \phi_{n+1,m-1} \right) + \mathrm{h.c.}\right)\\ \nonumber
\end{align}
It is easy to show that the only nonzero commutator relations are $[C,D]$, $[E,J]$, $[E,Y]$, and $[E,Z]$. Then we find
\begin{align}
[&[H_0,A_1],A_1] = \frac{\omega_{\mathrm{rec}}^2 (f_0\epsilon_0)^2}{16}\biggl[4\omega_{\mathrm{rec}}[[E,Y],Y] -\frac{2\sqrt{\omega_{\mathrm{rec}}}\sqrt{|\Delta_0|}}{\sqrt{\epsilon_0}}D \biggl([[E,J],Y]+[[E,Y],J]\biggr) \\ \nonumber
&-\frac{\Delta_0}{{\epsilon_0}\omega_{\mathrm{rec}}}[[C,D],D]J^2\left(-\delta_{\mathrm{C}}+\frac{\Delta_0}{2}\left(N+\frac{Z}{2}\right)\right) -\frac{\Delta_0}{\epsilon_0}D^2[[E,J],J]  \biggr].
\end{align}
One useful property for calculating commutators between various momentum mode operators is
\begin{align}
\sum_{n,m,n',m'} &[f(n,m)\phi^{\dagger}_{n+a,m+b}\phi_{n+c,m+d},\phi^{\dagger}_{n'+a',m'+b'}\phi_{n'+c',m'+d'}] \\ \nonumber
&=\sum_{n,m}\biggl(f(n,m)\phi^{\dagger}_{n+a,m+b}\phi_{n+c+c'-a',m+d+d'-b'}-f(n+c'-a,m+d'-b) \phi^{\dagger}_{n+a',m+b'}\phi_{n+c+c'-a,m+d+d'-b} \biggr)
\end{align}
Using this property, we get
\begin{align}
&[[H_0,A_1],A_1] = \frac{\omega_{\mathrm{rec}}^2 (f_0\epsilon_0)^2}{16}\biggl[32{\omega_{\mathrm{rec}}}\left( \sum(\phi^{\dagger}_{n,m}\phi_{n-4,m} + \mathrm{h.c.})\right) +\frac{2\Delta_0}{{\epsilon_0}\omega_{\mathrm{rec}}}J^2\left(-\delta_{\mathrm{C}}+\frac{\Delta_0}{2}\left(N+\frac{Z}{2}\right)\right)\\ \nonumber
&-\frac{4\Delta_0}{\epsilon_0}(\alpha^{\dagger}+\alpha)^2\left(-4\sum\phi^{\dagger}_{n,m}\phi_{n,m} + \sum(\phi^{\dagger}_{n,m}(\phi_{n+2,m-2}+\phi_{n+2,m+2}) + \mathrm{h.c.})  \right) \\ \nonumber
&-16\frac{\sqrt{\omega_{\mathrm{rec}}}\sqrt{|\Delta_0|}}{\sqrt{\epsilon_0}}(\alpha^{\dagger}+\alpha)\biggl(\sum(\phi^{\dagger}_{n,m}(\phi_{n+3,m-1}+\phi_{n+3,m+1}-(\phi_{n+1,m-1}+\phi_{n+1,m+1})) + \mathrm{h.c.}) \biggr) \biggr] 
\end{align}
Then the first nontrivial correction to the effective Hamiltonian reads
\begin{align}\label{eq:ham2}
&H^{(2)}_{\mathrm{eff}}=-\frac{\omega_{\mathrm{rec}}^3 (f_0\epsilon_0)^2}{2\omega_d^2}\left( \sum(\phi^{\dagger}_{n,m}\phi_{n-4,m} + \mathrm{h.c.})\right) - \frac{ \omega_{\mathrm{rec}}\Delta_0 (f_0\epsilon_0)^2}{32\omega_d^2\epsilon_0}\left(-\delta_{\mathrm{C}}+\frac{\Delta_0}{2}\left(N+\frac{Z}{2}\right)\right)J^2 \\ \nonumber
&+\frac{\sqrt{\omega_{\mathrm{rec}}}\sqrt{|\Delta_0|}\sqrt{\epsilon_0} }{4}\epsilon_0\left(\frac{\omega_{\mathrm{rec}}}{\omega_d}\right)^2(f_0)^2(\alpha^{\dagger}+\alpha)\sum(\phi^{\dagger}_{n,m}(\phi_{n+3,m-1}+\phi_{n+3,m+1})+ \mathrm{h.c.}) \\ \nonumber
&+\frac{\epsilon_0}{2}\frac{\Delta_0}{8}\left(\frac{\omega_{\mathrm{rec}}}{\omega_d}\right)^2(f_0)^2(\alpha^{\dagger}+\alpha)^2\sum(\phi^{\dagger}_{n,m}(\phi_{n+2,m-2}+\phi_{n+2,m+2}) + \mathrm{h.c.}) \\ \nonumber
&-\frac{\sqrt{\omega_{\mathrm{rec}}}\sqrt{|\Delta_0|}\sqrt{\epsilon_0} }{4}\epsilon_0\left(\frac{\omega_{\mathrm{rec}}}{\omega_d}\right)^2(f_0)^2(\alpha^{\dagger}+\alpha)\sum(\phi^{\dagger}_{n,m}(\phi_{n+1,m-1}+\phi_{n+1,m+1})+ \mathrm{h.c.}) \\ \nonumber
&-\frac{\epsilon_0}{2}\frac{\Delta_0}{2}\left(\frac{\omega_{\mathrm{rec}}}{\omega_d}\right)^2(f_0)^2(\alpha^{\dagger}+\alpha)^2\sum\phi^{\dagger}_{n,m}\phi_{n,m}
\end{align}
Note that the first two lines in the Eq.~\eqref{eq:ham2} can be neglected \textit{a posteriori}. In the first line, the first term can be dropped when higher momentum modes corresponding to $\{n+4,m+4\}$ for any any integer values of $n$ and $m$ have negligible occupation which is the case for all superradiant states obtained in this work as exemplified by the DW$_1$ state in Fig. 5(r). The second term, on the other hand, will have negligible contribution since $J \ll 1$ is almost zero for the BEC phase while it will be several orders of magnitude lower than the next relevant energy scale in the Hamiltonian for the self-organized phase. The second and third lines corresponding to higher-order hopping terms in momentum space can also be neglected for moderate depletion of the BEC mode such that $|\phi_{0,0}|^2+\sum_{n,m=\{\pm 1, \pm 1\}}|\phi_{n,m}|^2 \approx N_a$. This simplification is further justified in calculations considered here since we focus around the phase transition boundary where there are still relatively fewer photons occupying the cavity mode in the DW$_1$ phase. Finally, the effective time-independent Hamiltonian is given by
\begin{align}
&H_{\mathrm{eff}}=-\delta_{\mathrm{C}} {\alpha}^{\dagger} {\alpha} + \frac{\Delta_0}{4} {\alpha}^{\dagger} {\alpha}\sum_{n,m}\left(  {\phi}^{\dagger}_{n,m+2} {\phi}_{n,m} + \mathrm{h.c.} \right) + \omega_{\mathrm{rec}}\sum_{n,m}(n^2+m^2) {\phi}^{\dagger}_{n,m} {\phi}_{n,m}+\frac{\Delta_0}{2} {\alpha}^{\dagger} {\alpha}\sum_{n,m} {\phi}^{\dagger}_{n,m} {\phi}_{n,m} \\ \nonumber
&-\frac{\omega_{\mathrm{rec}} \epsilon_0}{2}\left[1+\frac{f_0^2}{2}+\frac{\Delta_0}{2\omega_{\mathrm{rec}}}\left(\frac{\omega_{\mathrm{rec}}}{\omega_d}\right)^2f_0^2(\alpha^{\dagger}+\alpha)^2\right] \sum_{n,m} {\phi}^{\dagger}_{n,m} {\phi}_{n,m} -\frac{\omega_{\mathrm{rec}} \epsilon_0}{4}\left(1+\frac{f_0^2}{2}\right)\sum_{n,m}\left(  {\phi}^{\dagger}_{n,m} {\phi}_{n+2,m} + \mathrm{h.c.} \right) \\ \nonumber
&+\frac{\sqrt{\omega_{\mathrm{rec}}} \sqrt{|\Delta_0|}\sqrt{\epsilon_0} }{4}\left[1-\epsilon_0\left(\frac{\omega_{\mathrm{rec}}}{\omega_d}\right)^2f_0^2 \right](\alpha^{\dagger}+\alpha)\sum(\phi^{\dagger}_{n,m}(\phi_{n+1,m-1}+\phi_{n+1,m+1})+ \mathrm{h.c.})
\end{align}
\begin{figure}[!htb]
\centering
\includegraphics[width=0.5\columnwidth]{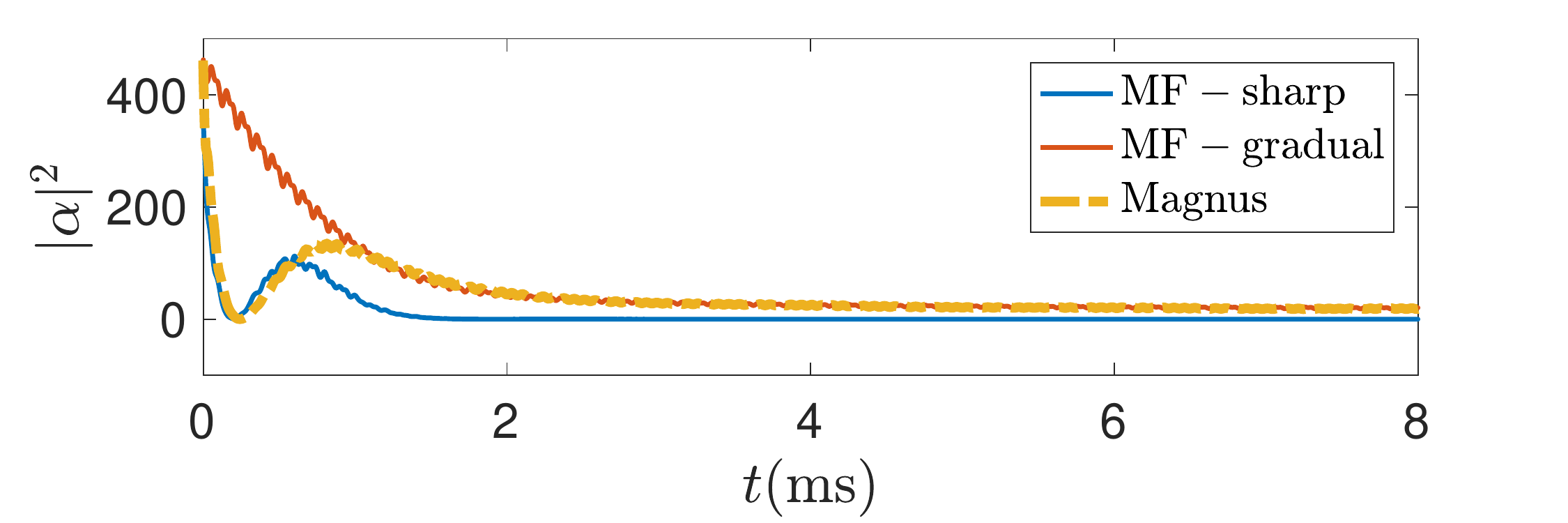}
\caption{Comparison of the cavity mode dynamics between the solution of the full mean-field equations and the effective time-independent Hamiltonian from the Magnus expansion. The driving frequency is set to $w_d=2\pi\times 10~\mathrm{kHz}$ and the driving amplitude is $f_0=0.152$. }
\label{fig:mag} 
\end{figure}
Upon normalization of the momentum mode occupation $\sum \phi^*_{n,m}\phi_{n,m} = 1$, we finally obtain the corresponding mean-field equation for the effective Hamiltonian $H_{\mathrm{eff}}$
\begin{align}\label{eq:eom_mag}
	&i \frac{\partial \phi_{n,m}}{\partial t} = 
	\omega_{\mathrm{rec}}\left(n^2+m^2+\frac{\Delta_0}{2\omega_{\mathrm{rec}}}|\alpha|^2-\frac{\epsilon_0}{2}\left(1+\frac{f_0^2}{2}+2(\mathrm{Re}(\alpha))^2\frac{\Delta_0}{\omega_{\mathrm{rec}}}\left(\frac{\omega_{\mathrm{rec}}}{\omega_d}\right)^2f_0^2 \right) \right)\phi_{n,m} \\ \nonumber
	 &+\frac{\Delta_0}{4}|\alpha|^2(\phi_{n,m-2}+\phi_{n,m+2})-\frac{\omega_{\mathrm{rec}}\epsilon_0}{4}\left(1+\frac{f_0^2}{2}\right)(\phi_{n-2,m}+\phi_{n+2,m})\\ \nonumber
	 &+\frac{\sqrt{\omega_\mathrm{rec}}\sqrt{|\Delta_0|}\sqrt{\epsilon_0}}{2}\left[1-\epsilon_0\left(\frac{\omega_{\mathrm{rec}}}{\omega_d}\right)^2f_0^2 \right]\mathrm{Re}(\alpha)(\phi_{n-1,m-1}+\phi_{n+1,m-1}+\phi_{n-1,m+1}+\phi_{n+1,m+1}) \\ \nonumber
	 &i \frac{\partial \alpha}{\partial t} = \left(\left(-\delta_{\mathrm{eff}} -\frac{N_a\Delta_0 \epsilon_0}{2}\left(\frac{\omega_{\mathrm{rec}}}{\omega_d}\right)^2f_0^2 \right) +\frac{1}{2}N_a\Delta_0\sum_{n,m}\mathrm{Re}[\phi_{n,m}\phi^{*}_{n,m+2}]-i\kappa \right)\alpha \\ \nonumber
	 &+\frac{N_a\sqrt{\omega_{\mathrm{rec}}}\sqrt{|\Delta_0|}}{4}\sqrt{\epsilon_0}\left[1-\epsilon_0\left(\frac{\omega_{\mathrm{rec}}}{\omega_d}\right)^2f_0^2 \right]\sum_{n,m}\left(\phi_{n,m}(\phi^{*}_{n+1,m+1}+\phi^{*}_{n+1,m-1})+ \mathrm{h.c.}\right)+\frac{N_a\Delta_0 \epsilon_0}{2}\left(\frac{\omega_{\mathrm{rec}}}{\omega_d}\right)^2f_0^2\alpha^*.
\end{align}
Note that we recover the  mean-field equations of motion in Ref.~\cite{Klinder2015} for the undriven case $f_0=0$. From the bracketed terms in Eq.~\ref{eq:eom_mag}, it is easy to see that the enhancement of the BEC phase can be explained by an effective reduction in the coupling strength of the two-photon process that scatters atom from $\phi_{0,0}$ to $\phi_{\pm 1, \pm 1}$ 
\begin{equation}
\sqrt{\epsilon_0}  \xrightarrow{\text{driven}} \sqrt{\epsilon_0}\left[1-\epsilon_0\left(\frac{\omega_{\mathrm{rec}}}{\omega_d}\right)^2f_0^2 \right].
\end{equation}

We numerically integrate this set of equations in order to obtain the results shown in thin solid lines in Fig. 4. We also show in Fig.~\ref{fig:mag} a comparison between the results of numerically integrating the full mean-field equations and those from an effective time-independent Hamiltonian according to Eq.~\eqref{eq:eom_mag}. 
\begin{figure}[!ht]
\centering
\includegraphics[width=0.4\columnwidth]{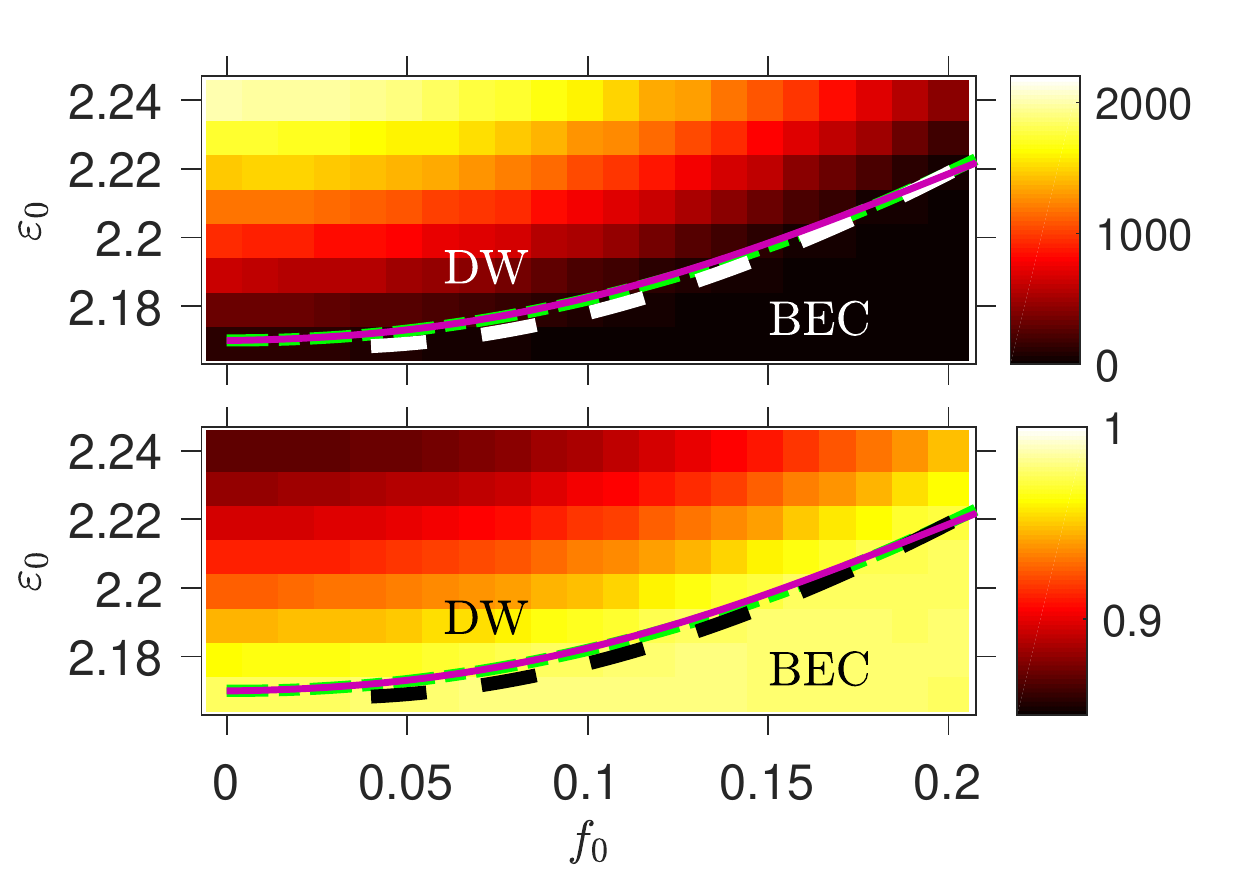}
\vspace{-0.4cm}
\caption{Dynamical renormalization of the BEC-DW phase transition, visible in the (top) cavity mode  and (bottom) BEC mode occupation for $w_d=2\pi\times 10~\mathrm{kHz}$. (i) Thin solid line shows the effective Hamiltonian prediction for the phase boundary, (ii) thick dashed line the TW result, (ii) thick dashed-dotted line the MF result. The phase boundary is indicated based on $|\alpha|^2> 70$ and $n_0/N> 0.97$.}
\label{fig:mf_tw_phase} 
\end{figure}
For the driven case presented in Fig.~\ref{fig:mag}, we have applied a Gaussian filter with width $\sigma=1/\omega_{\mathrm{rec}}$ to artificially remove the micromotion part of the dynamics which is inherently not captured by the effective Hamiltonian obtained here. In doing so, we can then focus on more important aspects of the dynamics including its overall trend and long-time behaviour. On one hand, we find that the effective time-independent Hamiltonian nicely captures the short time dynamics predicted by the full mean-field equations after the modulation is sharply switched on. This suggests that the driving protocol can be seen as some kind of sudden quench to an effectively weaker atom-cavity coupling. On the other hand, we find that the steady-state predictions from the effective Hamiltonian agree very well with the mean-field counterpart for a gradual ramp of the driving amplitude as exemplified in Fig.~\ref{fig:mag}. This is of course consistent with the good agreement for the phase boundary shown in Fig.~\ref{fig:mf_tw_phase}.

\section{Mean-field order parameters and single-particle density profiles for density-wave ordered phases}

\begin{figure}[!ht]
\centering
\includegraphics[width=0.33\columnwidth]{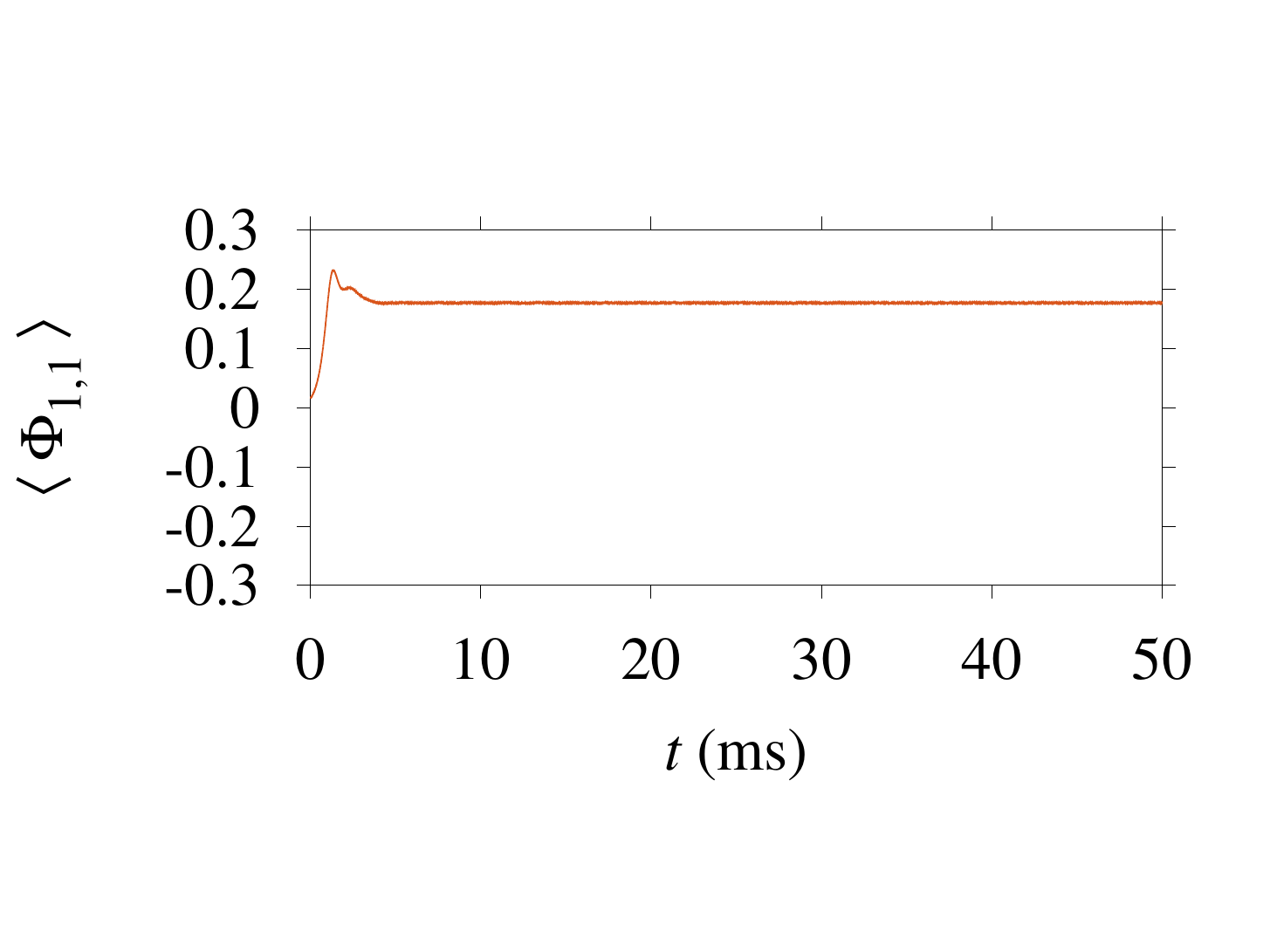}\includegraphics[width=0.33\columnwidth]{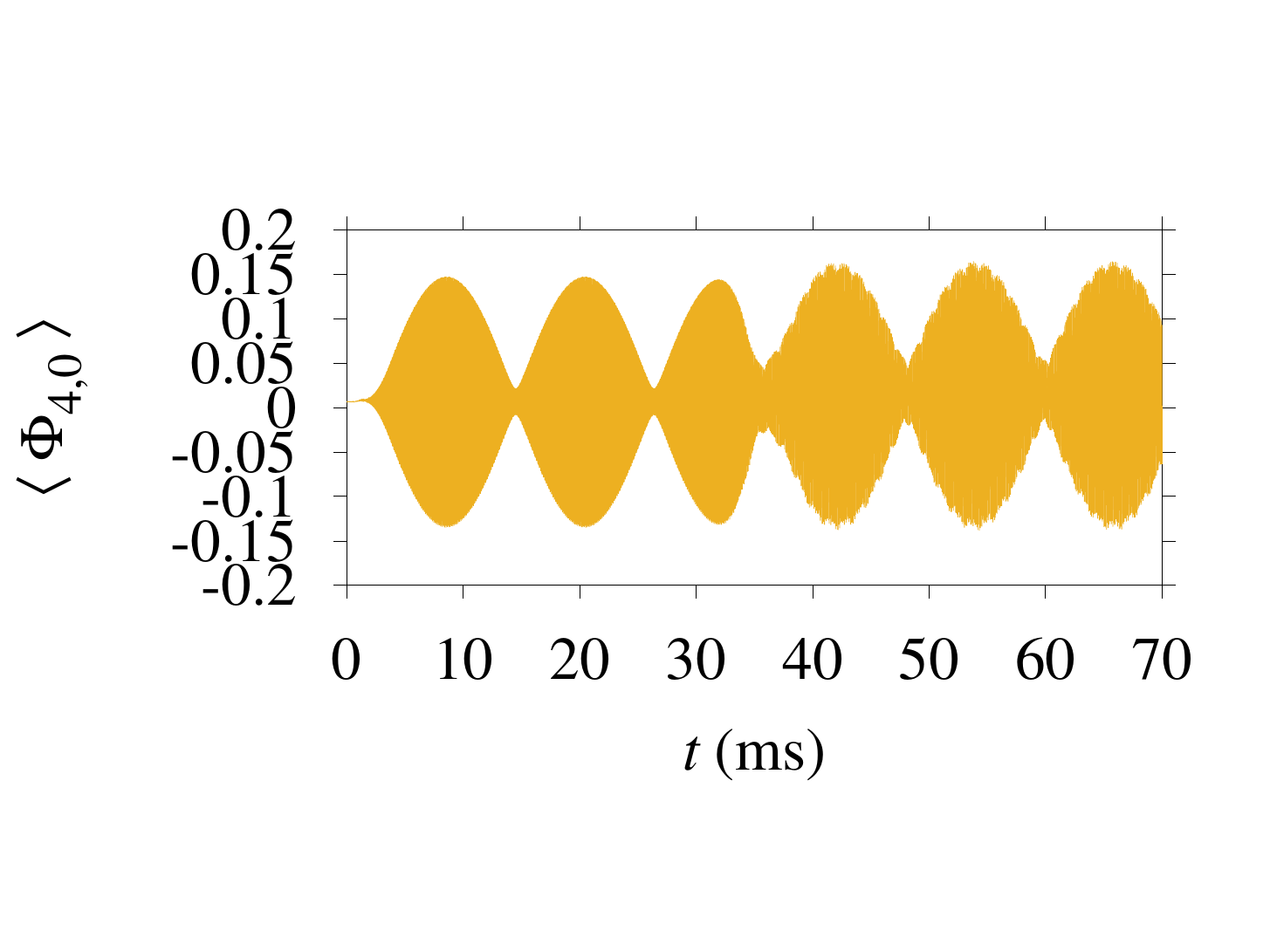}\includegraphics[width=0.33\columnwidth]{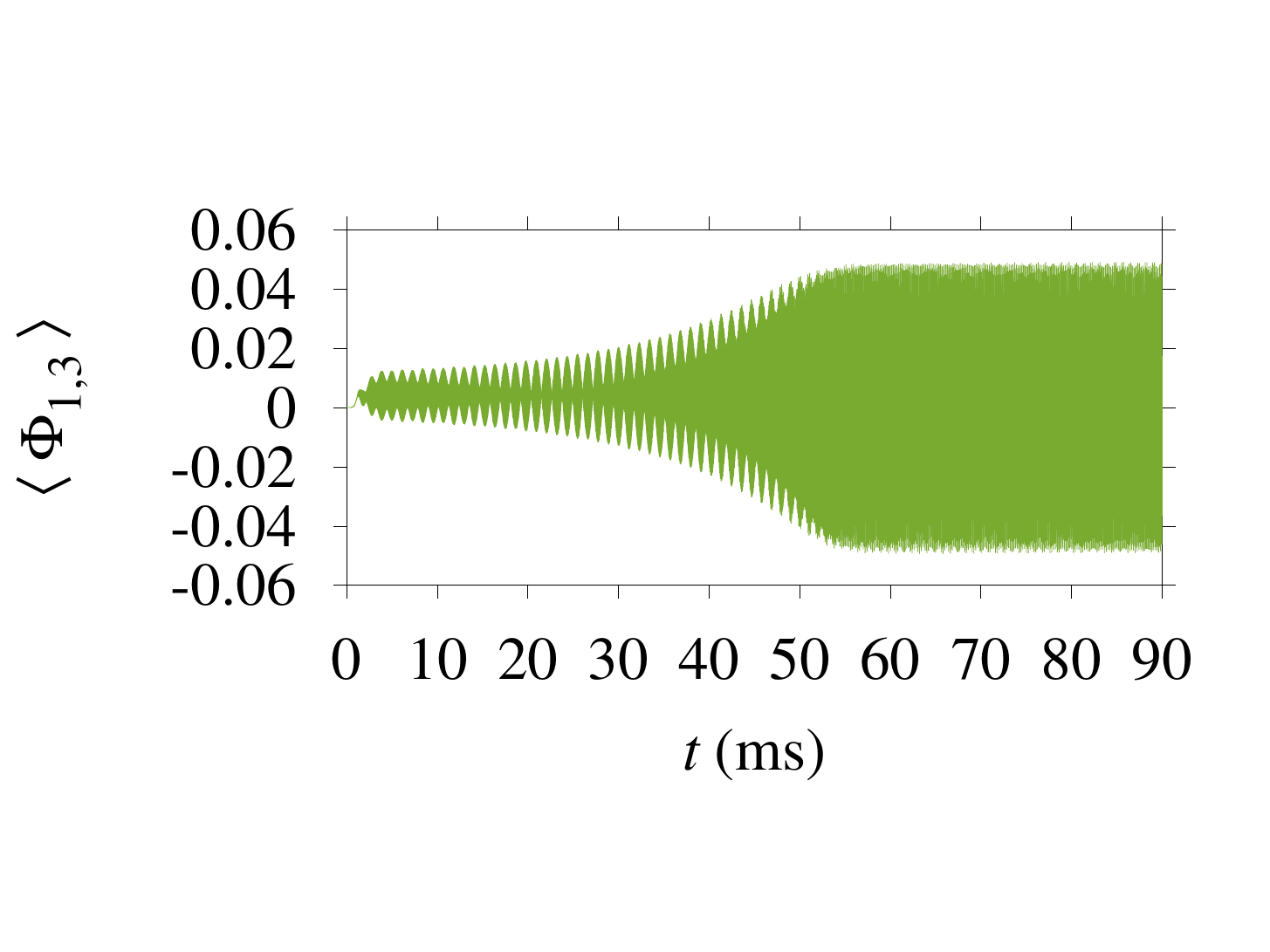}\\
\vspace{-1cm} 
\hspace{-0.5cm}\includegraphics[trim={0 0 4cm 0},clip,width=0.25\columnwidth]{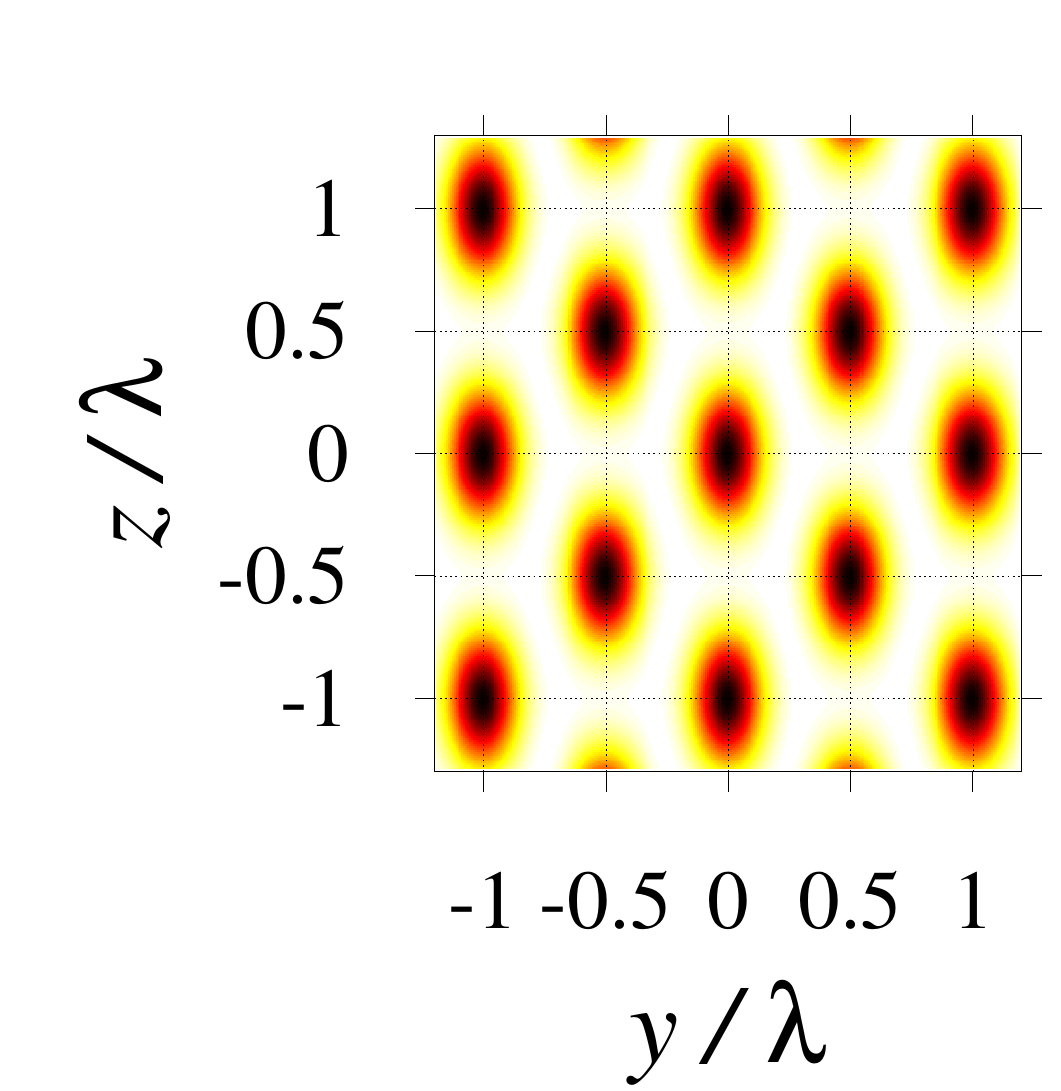}\hspace{1.5cm}\includegraphics[trim={0 0 4cm 0},clip,width=0.25\columnwidth]{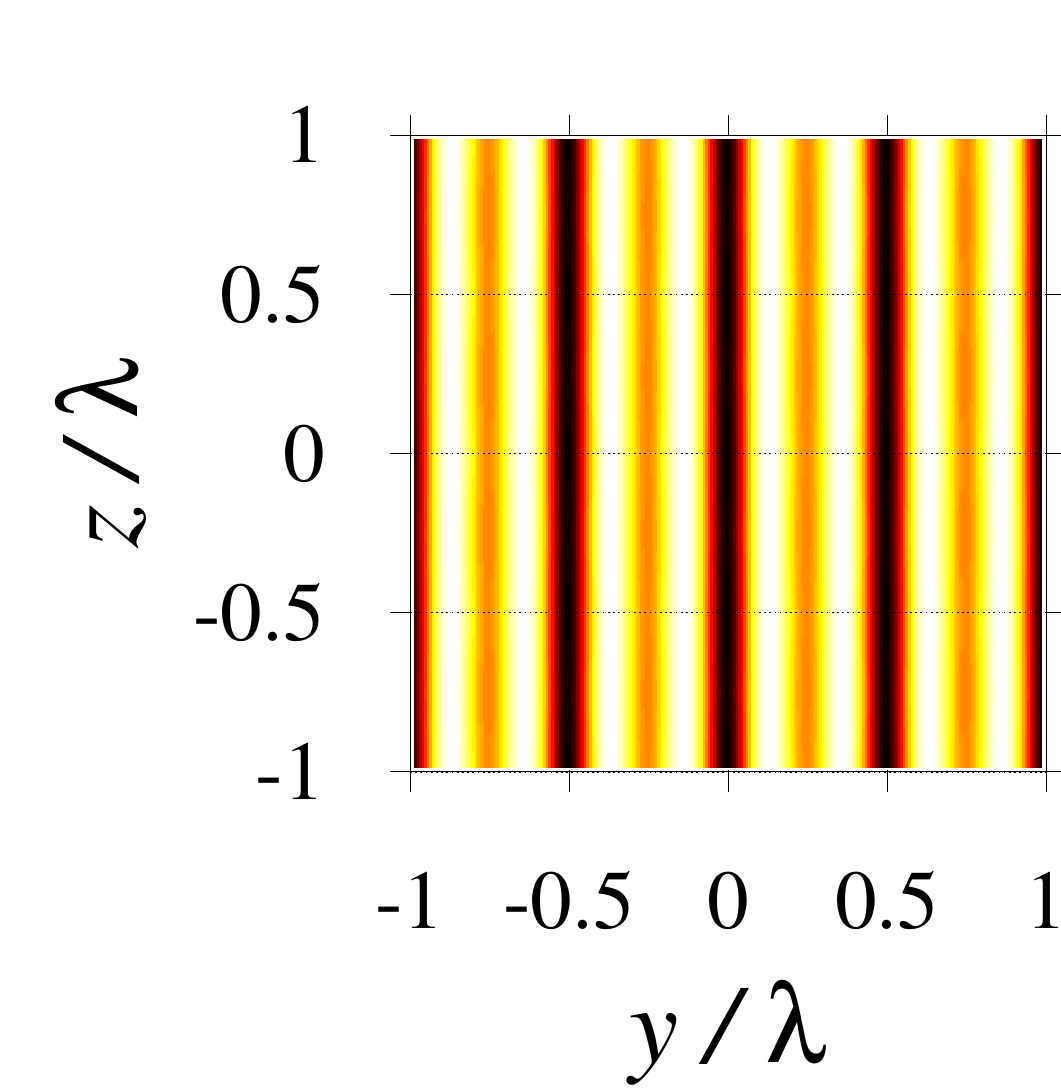}\hspace{1.5cm}\includegraphics[trim={0 0 4cm 0},clip,width=0.25\columnwidth]{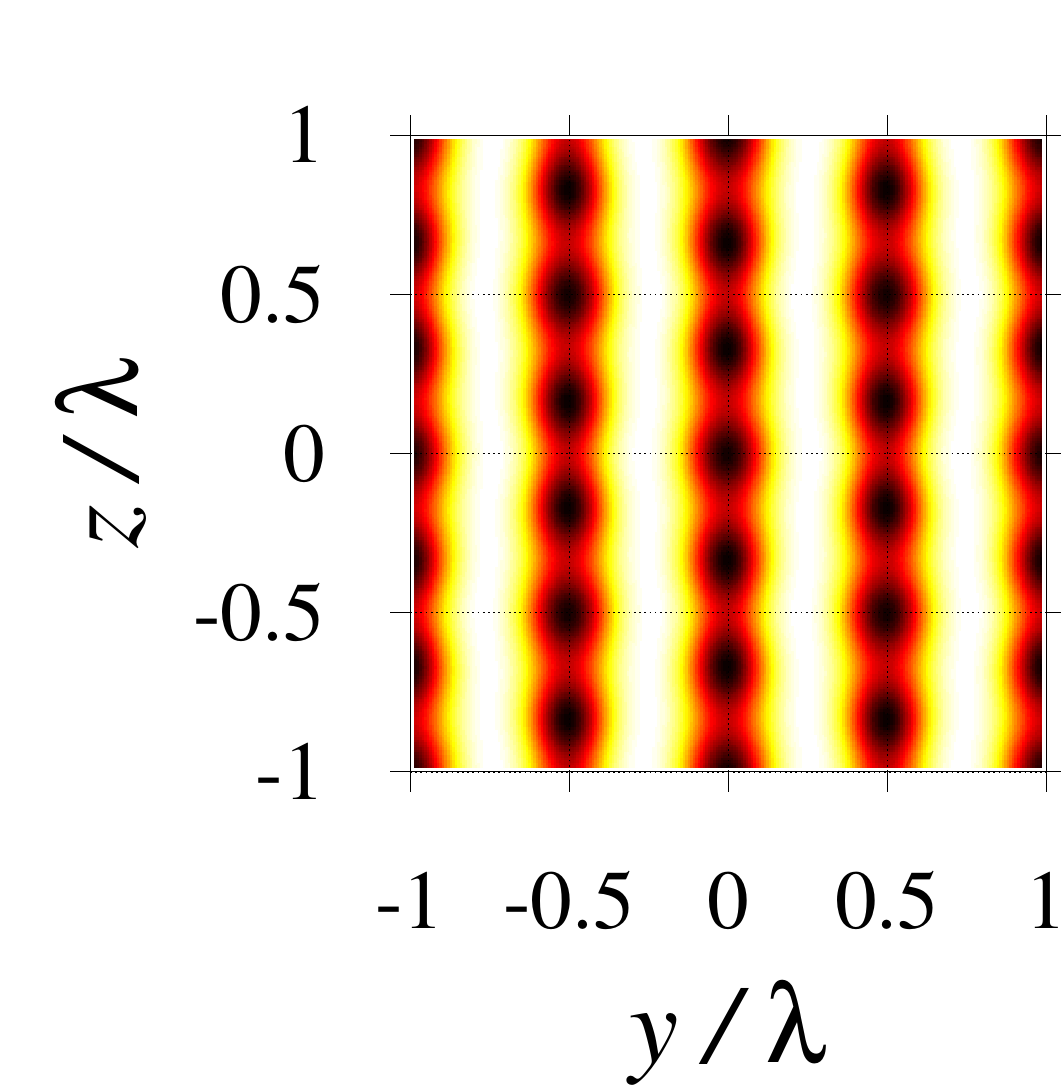}
\caption{(Top) Dominant order parameters and (Bottom) exemplary single-particle density profiles within the mean-field theory for (left) DW$_1$ ($w_d=2\pi\times 34.0~\mathrm{kHz}$), (middle) DW$_4$ ($w_d=2\pi\times 28.5~\mathrm{kHz}$), and (right) DW$_3$ ($w_d=2\pi\times 35.5~\mathrm{kHz}$). The exact parameters are the same as Fig. 5 in the main text.}
\label{fig:mf_op} 
\end{figure}

Here, we present results for single trajectories in our TW simulations, which basically correspond to mean-field predictions for the dynamics. In particular, we calculate the expectation value of the dominant order parameter $\langle \Phi_{n,m} \rangle$ for the DW$_1$,  DW$_4$, and  DW$_3$ dynamical phases. We also obtain exemplary single-particle density (spd) profiles, $\rho(y,z)= \sum_{n,m,n',m'} \phi^{\dagger}_{n,m}\phi_{n',m'}e^{i(n-n')ky}e^{i(m-m')ky}$, in the long-time limit of each DW phases in order to gain further insights on possible symmetry breaking phenomenon. The corresponding results are shown in Fig.~\ref{fig:mf_op}. 

For the renormalized DW$_1$ phase in the presence of driving, the original $\mathbb{Z}_2$-symmetry breaking associated to the self-organization of atoms survives as seen in the left panel of Fig.~\ref{fig:mf_op}. In this case, the atoms spontaneously form one of the two possible checkerboard patterns corresponding to a positive-valued order parameter $\langle \Phi_{1,1} \rangle$. Moreover, the small temporal fluctuation of the leading order parameter suggests that the atomic ensemble essentially remains fixed in one of the symmetry broken ordered phases for long times.

In contrast to the nonequilibrium DW$_1$ phase, we find that the DW$_4$ and DW$_3$ phases exhibit strong oscillation of the dominant order parameters around zero. This physically means that the system is dynamically switching between possible symmetry broken ordered phases. This phenomenon has been also predicted for the so-called dynamical normal phase \cite{Chitra2015,Molignini2017} where the atoms are dynamically switching between the even and odd checkerboard patterns. For the DW$_4$ phase shown in the middle panel of Fig.~\ref{fig:mf_op}, the system is oscillating between possible striped phases corresponding to density modulation along the direction of the pump beam. This is consistent with the absence of momentum excitations along the $z$-direction shown in Fig. 5(q). Similarly, the single-particle density profile for the DW$_3$ phase dynamically switches between stripe-ordered phases with additional checkerboard density modulation along the cavity axis as depicted in Fig.~\ref{fig:mf_op}.

\section{Importance of the recoil resolution $\kappa$}

\begin{figure}[!ht]
\centering
\includegraphics[width=0.5\columnwidth]{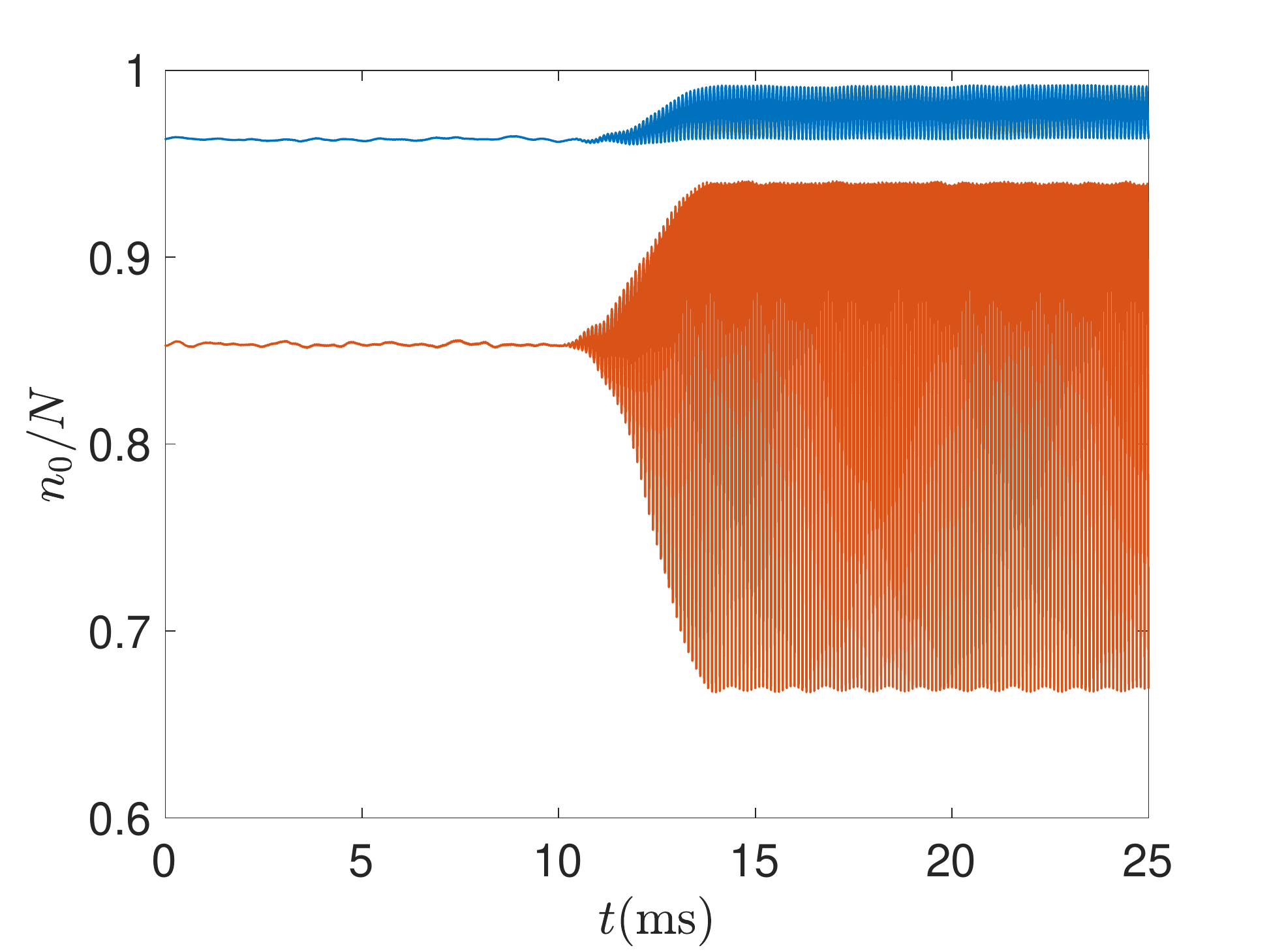}\includegraphics[width=0.5\columnwidth]{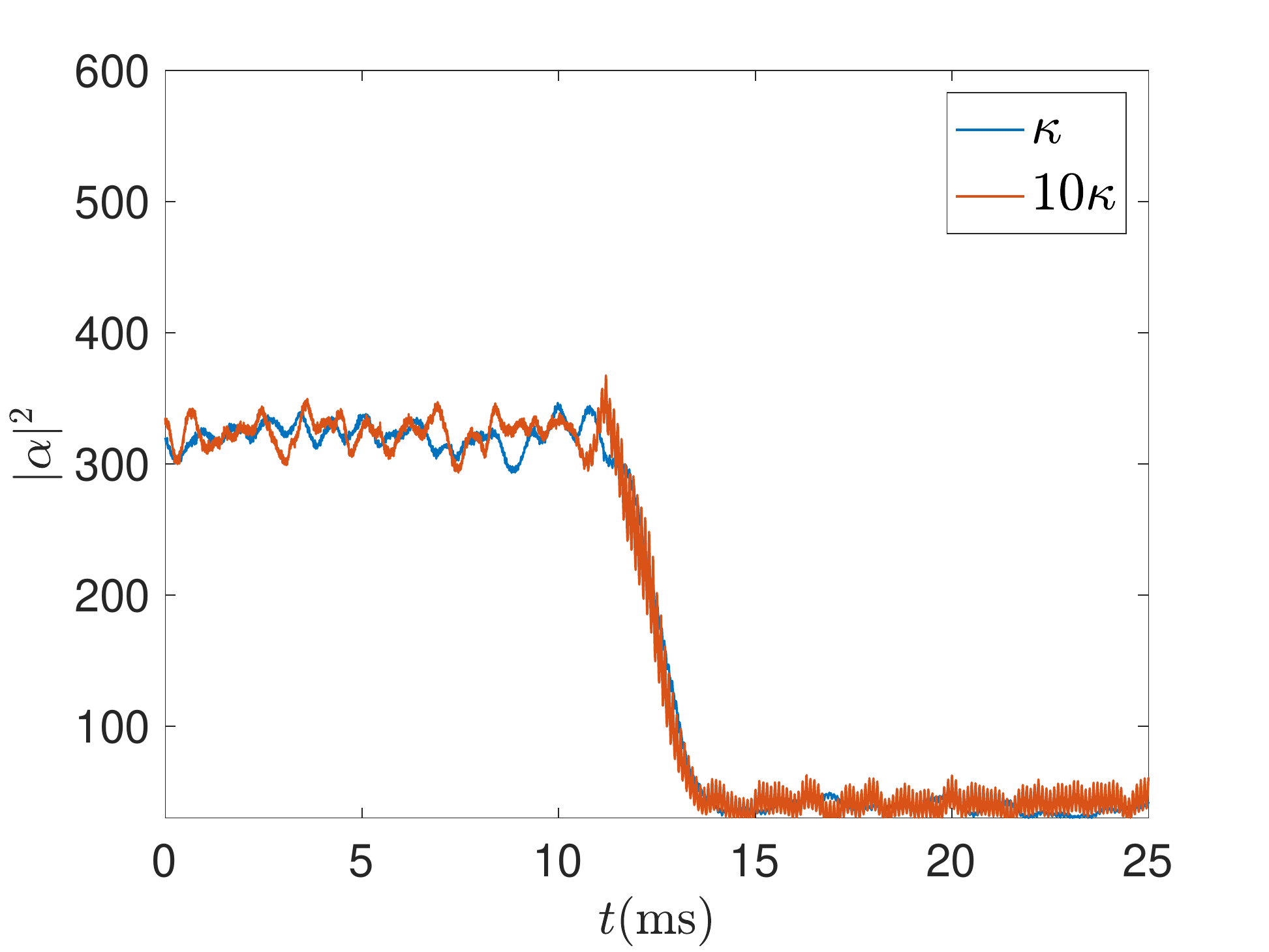}
\caption{Comparison of the (left) BEC and (right) cavity modes for $\kappa=\kappa_{\mathrm{expt}}$ ($\kappa=10\kappa_{\mathrm{expt}}$) with $f_0=0.12~(0.22)$, $\varepsilon_0/E_{\mathrm{rec}}=2.18~(6.93)$, and $w_d=2\pi\times 10~\mathrm{kHz}$.}
\label{fig:tw_kappa} 
\end{figure}

In the case when $4\omega_{\mathrm{rec}} \ll \kappa$ just like in Ref.~\cite{Baumann2010}, the cavity mode adiabatically follows the atomic degrees of freedom such that only the dynamics of the atomic modes need to be considered explicitly. 
As mentioned in the main text, we find that it is important to explicitly consider the dynamics of both the atomic and cavity modes in order to mimic the dynamical suppression effect of density-wave order seen in high-$T_c$ superconductors. That is, we briefly show here the importance of having  $4\omega_{\mathrm{rec}} \gg \kappa$ as in Refs.~\cite{Klinder2015,Klinder2015b,Klinder2016} in the recondensation process after the modulation. To this end, we show in Fig.~\ref{fig:tw_kappa} the ensuing dynamics for the BEC and cavity modes for $\kappa=\kappa_{\mathrm{expt}}=2\pi\times 4.5~\mathrm{kHz}$ and for $\kappa=10\kappa_{\mathrm{expt}}$. We adjust the mean pump strength for each case in order to fix the number of photons in the DW phase. We then choose a critical modulation amplitude $f_0$ which is just enough to completely suppress the cavity mode occupation. A higher value for $\kappa$ means that the mean pump strength needed to enter the DW phase will have to increase as well as evident from our simulation. Even though we are still able to completely suppress the DW phase for $\kappa=10\kappa_{\mathrm{expt}}$, the number of atoms that we recover back to the BEC mode is not significant in contrast to the case when $\kappa=\kappa_{\mathrm{expt}}$. Moreover, we find stronger temporal variance in the BEC mode occupation for $\kappa=10\kappa_{\mathrm{expt}}$. These observations suggest the importance of low $\kappa$ and it also emphasizes the point that the photonic and atomic degrees of freedom should be treated individually.

\end{document}